\documentclass[reprint,amsmath,amssymb,aps,prx,floatfix]{revtex4-2}

\usepackage{graphicx}
\usepackage{dcolumn}
\usepackage{bm}
\usepackage{xcolor}
\usepackage{booktabs}

\newcommand{\eps}{\epsilon}
\newcommand{\ee}{\varepsilon}

\newcommand{\im}{\mathrm{Im}}
\newcommand{\re}{\mathrm{Re}}

\newcommand{\kB}{k_{\rm B}}
\newcommand{\vF}{v_{\rm F}}
\newcommand{\nuF}{\nu_{\rm F}}
\newcommand{\dd}{\mathrm{d}}
\newcommand{\ii}{\mathrm{i}}
\newcommand{\thf}{\mathrm{t}}

\begin{document}

\preprint{APS/123-QED}

\title{Evanescent-wave Johnson Noise from Superconductors}
\author{Hruday Mallubhotla}
\author{Gustav Romare}
\author{Ilya Esterlis}
\author{Maxim Vavilov}
\author{Robert Joynt}
\author{Alex Levchenko}
\affiliation{Department of Physics, University of Wisconsin-Madison, Madison, Wisconsin 53706, USA}

\date{September 28, 2026}

\begin{abstract}
We compute the evanescent-wave Johnson noise (EWJN) in the vacuum half-space above a superconductor, and the resulting relaxation time ($T_1$) of spin and charge qubits placed at nanometer distances from the surface. The electromagnetic response is described by a single microscopic transverse current-response kernel $Q(q, \omega)$ for a BCS superconductor. This is computed for varying densities of both non-magnetic impurities 
and magnetic impurities, for arbitrary frequency and temperature and for wave vectors $q \ll k_F$ (the Fermi wavevector). When combined with the fluctuation-dissipation theorem and the nonlocal surface impedances of the half-space, this yields the magnetic and electric field noise at any distance $z \gg k_F^{-1}$ from the surface, from which we obtain $T_1$. Just below $T_c$ the magnetic noise is enhanced relative to the normal state by the coherence (Hebel-Slichter-type) peak of the dissipative conductivity and drops exponentially at lower temperatures; the electric noise shows no coherence peak. The theory predicts that there is a zero-temperature noise floor induced by magnetic impurities. In the gapless regime produced by pair breaking, the finite subgap density of states $\nu(0)$ yields a temperature-independent noise spectral density and a relaxation rate bounded by $T_1^{-1}(T)\le[\nu(0)/\nu_F]^{2}\,T_{1,N}^{-1}(T)$ for $T\ll T_c$, with equality in the extreme nonlocal regime. Here $\nu(0)$ and $\nu_F$ are the superconducting and normal-state densities of states at the Fermi energy, and $T_{1,N}(T)$ is the relaxation time the same electrode would produce in its normal state at the same temperature.
\end{abstract}

\maketitle

\section{Introduction}
\label{sec:intro}

The thermal and quantum agitations of charges in a conductor produce fluctuating electromagnetic fields. Inside a circuit, these fluctuations are observed as Johnson-Nyquist noise; in the space surrounding the conductor they appear as a fluctuating near field. At distances $z$ from a metallic surface small compared with the photon wavelength $\lambda = c/\omega$, the field noise is dominated not by propagating photons but by the evanescent components that leak out of the overdamped electromagnetic modes of the metal. The noise power at such distances exceeds the black-body value by many orders of magnitude \cite{loomis1994}. This evanescent-wave Johnson noise (EWJN) was first described in the framework of fluctuational electrodynamics by Rytov and Lifshitz \cite{rytov,lifshitz}, and its general machinery, the fluctuation-dissipation theorem (FDT) applied to the electromagnetic Green's function of the geometry, underlies the theory of the van der Waals/Casimir force and of near-field radiative heat transfer \cite{joulain}.

For quantum devices, the practical importance of EWJN is that it relaxes and decoheres qubits. Atomic qubits near metallic trap electrodes were the first studied cases \cite{henkel}. For solid-state spin and charge qubits near metallic gates the theory was developed in Refs.~\cite{langsjoen2012,poudel2013,langsjoen2014}. These works established that (1) the relaxation rate follows from the FDT and the reflection coefficients of the surface; (2) at distances $z$ smaller than the electromagnetic coherence length of the metal (mean free path $\ell$, or $v_F/\omega$ where $\omega$ is the qubit frequency), a nonlocal description of the dielectric response is required; and (3) the noise magnitude can serve as an experimental probe of the nonlocal conductivity $\sigma(q,\omega)$ of the electrode material. Indeed, measurements of the relaxation rates of nitrogen-vacancy (NV) centers placed a few tens of nanometers above silver films \cite{kolkowitz2015} confirmed that nonlocality is qualitatively important in understanding the noise strength of normal metals. 

More recently, EWJN has been viewed from the opposite direction as a tool rather than as a nuisance. Analysis of the magnetic noise gives information about the full nonlocal dielectric function in the microwave regime with good spatial resolution \cite{ariyaratne2018}. It can also serve as a probe for magnon dynamics \cite{vandersar2015,chatterjee2019,purser2020}. NV relaxometry has also been used to probe current noise characteristics of various materials \cite{casola2018probing,rovny2024,thiel2019probing,bhattacharyya2024,ku2020,du2017,andersen2019}, most recently including metals cooled below their superconducting transition temperature $T_c$ \cite{liu2025,li2026}. There is also the possibility that the noise is sensitive to the nature of the order parameter in two-dimensional  superconductors \cite{chatterjee2022,dolgirev2022, curtis2024,cheng2026}.
EWJN is correlated in space and time \cite{premakumar2018}. This can be used to characterize the spatial and temporal structure of electromagnetism of solid state systems in even more detail \cite{rovny2022,le2025,rovny2025,cheng2025,huxter2025,cambria2025,hosseinabadi2026,romare2026,orgad2026}, both in and out of equilibrium. 

The theory of EWJN above a clean two-dimensional BCS superconductor, including the coherence-peak enhancement of the noise below $T_c$, has been worked out recently in Ref.~\cite{kelly2025} (see also Refs.~\cite{chatterjee2022,dolgirev2022} for unconventional order parameters). In the present work we treat a bulk (three-dimensional) superconductor and give the results for arbitrary concentrations of non-magnetic and magnetic impurities. The effects of disorder are important for the height and temperature dependence of the noise, even at zero temperature. We highlight three of our central results: First, below $T_c$ the magnetic noise falls off exponentially with decreasing temperature (a noise ``cliff''), so that superconducting electrodes can improve $T_1$ by many orders of magnitude. Second, the Hebel-Slichter-type coherence peak of the dissipative conductivity, which is prominent in the local conductivity $\sigma_1(T)$ just below $T_c$, survives in the noise at typical NV working distances: for niobium films the noise just below $T_c$ exceeds the normal-state value by a factor $\approx2$ at heights of $3$--$10$~nm, the excess being progressively reduced at larger heights by nonlocality and superfluid screening, which scales as the inverse square of the penetration depth. Third, magnetic impurities terminate the cliff: in the gapless regime the subgap density of states produces a temperature-independent noise floor, the qubit-relaxation analog of the residual surface resistance of superconducting radio-frequency (SRF) cavities. We discuss these effects in detail below.

This paper is organized as follows. In Sec.~\ref{sec:formalism}, we present the formalism: the relation between qubit relaxation and field correlation functions, the electromagnetic Green's functions of a half-space, and nonlocal surface impedances. We focus on magnetic noise, which is most relevant for NV sensing applications.  In Sec.~\ref{sec:results}, we present our main results for the dielectric function of a BCS superconductor as a function of $T$, $q$, and $\omega$, both in the clean limit as well as with non-magnetic and magnetic impurities. We use this to calculate the magnetic noise and spin-qubit relaxation time $T_1(T,z)$. The end of Sec.~\ref{sec:results} also includes a brief discussion of electric noise and charge qubit relaxation. Our conclusions are provided in Sec.~\ref{sec:conclusion}, which also contains a discussion of the experimental implications of our work. Details of the analytical and numerical calculations are contained in the Appendices.

\section{Formalism}
\label{sec:formalism}

In this section we summarize our formalism, starting first with the qubit relaxation rate due to fluctuations of a noisy electromagnetic field. We then present formulas relating the electromagnetic field noise to the reflection coefficients from a proximate conducting material, and relate the reflection coefficients to the nonlocal dielectric function of the material. Many of the derivations are standard and we have relegated the details to Appendices~\ref{app:half_space_greens_func} and \ref{app:nonlocal_impedances}.

\begin{figure}[t]
\centering
\includegraphics[width=\columnwidth]{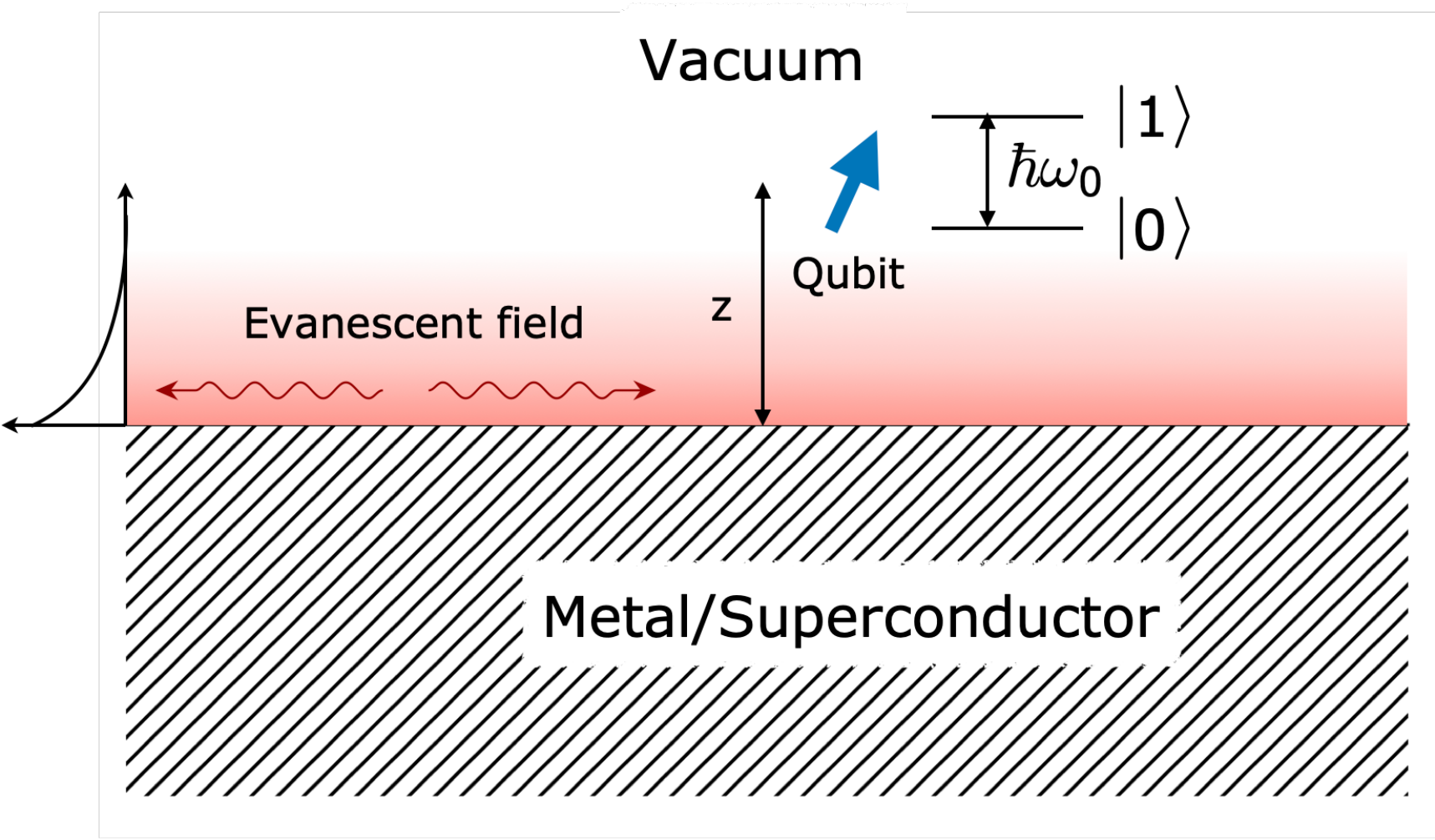}
\caption{A qubit with level splitting $\hbar\omega_0$ in vacuum, placed a distance $z$ above a metal which becomes a superconductor at low temperatures. Current fluctuations in the metal/superconductor generate evanescent electromagnetic waves that relax the qubit with the rate $1/T_1$.}
\label{fig:schem}
\end{figure}

\subsection{Qubit relaxation from field fluctuations}

Consider a qubit placed at a position $\mathbf r_0$ in vacuum above a metal. The set up is shown schematically in Fig.~\ref{fig:schem}. We describe the qubit as a two-level system with level splitting $\hbar \omega_0$ and eigenstates $|0\rangle$ and $|1\rangle$. A spin qubit couples to the magnetic field through $V = -\hat{\boldsymbol \mu} \cdot \mathbf B(\mathbf r_0)$, and a charge qubit couples to the electric field through $V = -\hat{\mathbf d} \cdot \mathbf E(\mathbf r_0)$. To second-order in $V$, Fermi's golden rule gives the transition rates
    \begin{equation}
        \Gamma_\downarrow  = \frac{1}{\hbar^2} \sum_{ij} \mu_i \mu_j^* S_{ij}(\omega_0), \quad \Gamma_\uparrow  = \frac{1}{\hbar^2} \sum_{ij} \mu_i \mu_j^* S_{ij}(-\omega_0),
    \end{equation}
where $\mu_{i} = \langle 0 | \hat \mu_i | 1 \rangle$ and
    \begin{equation}
        S_{ij}(\omega) = \int_{-\infty}^\infty {\rm d} t ~ e^{i\omega t} \langle B_i(\mathbf r_0, t) B_j(\mathbf r_0,0)\rangle
        \label{eq:sw}
    \end{equation}
is the (unsymmetrized) power spectral function of the field, with the average taken over the equilibrium state of the electromagnetic environment at temperature $T$. From the rate equations for the populations of the $|0\rangle$ and $|1\rangle$ states, the relaxation rate of the population difference is found to be $1/T_1 = \Gamma_{\downarrow} + \Gamma_{\uparrow}$.

Next, using the fluctuation dissipation theorem (FDT) we can relate the power spectral density \eqref{eq:sw} to the Green's function of the field: Consider a classical magnetic dipole $\mathbf m e^{-i\omega t}$ at position $\mathbf r'$. This dipole produces a field $B_i(\mathbf r) = G^B_{ij}(\mathbf r, \mathbf r', \omega)m_j$ where $G^B$ is the retarded response function of the field. For $\omega > 0$ the FDT gives    
\begin{subequations}
\begin{align}
        S_{ij}(\omega) &= 2\hbar \, [n_B(\omega) + 1]{\rm Im} G_{ij}^B(\omega), \\
        S_{ij}(-\omega) &= 2\hbar \, n_B(\omega){\rm Im}G_{ij}^B(\omega),
    \end{align}
\end{subequations}    
where $n_B(\omega) = (e^{\hbar \omega/ k_B T}-1)^{-1}$ is the Bose occupation factor. This leads to
    \begin{equation}
        \frac{1}{T_1} = \frac{1}{\hbar^2} \sum_{ij}\mu_i \mu_j^* \chi_{ij}^B(\mathbf r_0, \omega_0) \coth\left(\frac{\hbar \omega_0}{2k_B T}\right),
    \label{eq:T1master}
    \end{equation}
where $\chi_{ij}^B(\mathbf r_0, \omega) = 2\hbar \, {\rm Im}G^B_{ij}(\mathbf r_0, \mathbf r_0, \omega)$. The corresponding formulas for a charge qubit are obtained by the substitutions $B \to E$ and $\mu \to d$. This formula reproduces two well-known limits: in vacuum, where ${\rm Im}\,G^{0}_{ii}=\tfrac23(\omega/c)^3$, Eq.~\eqref{eq:T1master} at $T=0$ reproduces the spontaneous-emission rate $4\mu^2\omega^3/3\hbar c^3$ of a magnetic dipole; and for $k_BT\gg\hbar\omega_0$ the factor $\coth\to2k_BT/\hbar\omega_0$ reproduces the classical (Johnson) form of the FDT.

Let $\hat z$ be the surface normal. For a spin quantized along the $\hat n$ direction, the transverse operators drive the relaxation. In the special case where $\hat n$ is along $\hat z$ and the spin-qubit has spin-1/2 with $g$-factor $g=2$,  $1/T_1$ simplifies to
    \begin{equation}
        \frac{1}{T_1} = \frac{2\mu_B^2}{\hbar^2}\chi_\parallel^B(\mathbf r_0, \omega_0) \coth\left(\frac{\hbar \omega_0}{2k_BT}\right).
    \label{eq:T1_chi}
    \end{equation}
Here we assume a system isotropic about $\hat z$, so that $\chi_{xx} = \chi_{yy} \equiv \chi_\parallel$ and $\chi_{ij} = 0$ for $i \neq j$. A spin quantized in the plane would instead couple to  $\chi_\parallel + \chi_\perp$, where $\chi_\perp \equiv \chi_{zz}$. 

\subsection{Reflection coefficients}

The response function $\chi^{B}$ is related to the dielectric properties of the metallic system through the reflection coefficients $r_s$ and $r_p$ for $s$ (TE) and $p$ (TM) polarized waves, respectively. We define the dimensionless magnetic noise strength $b_{\parallel/\perp}^B(z,\omega) \equiv (4\hbar \omega^3 / 3 c^3)^{-1} \chi_{\parallel/\perp} ^B(z,\omega)$. Then
    \begin{subequations}
    \begin{align}
        b_\parallel^B(z,\omega) &= \frac 34 \int_0^\infty {\rm d}w ~ e^{-2kz w} \nonumber \\
        &\qquad \times {\rm Im}\left[r_p\left(\sqrt{1+w^2}\right) + w^2 r_s\left(\sqrt{1+w^2}\right)\right],
        \label{eq:bpar} \\
        b_\perp^B(z,\omega) &= \frac 32 \int_0^\infty {\rm d}w ~ (1+w^2) e^{-2 k z w}\,{\rm Im}\, r_s\left(\sqrt{1+w^2}\right),
        \label{eq:bperp}
    \end{align}
    \end{subequations}
where $k=\omega/c$, $u=cQ/\omega$ is the dimensionless lateral wave vector of the partial wave, and $w=\sqrt{u^2-1}$ is its (dimensionless) decay constant in the vacuum.
This result is valid in the near field where vacuum fluctuations and propagating ($u<1$) fields can be neglected, i.e., for $z\ll c/\omega$, and holds for arbitrary reflection coefficients. Equations  \eqref{eq:bpar} and \eqref{eq:bperp} are obtained from the half-space electromagnetic Green's function by standard methods; details are given in Appendix \ref{app:half_space_greens_func}. 

For the spin qubit, combining Eqs.~\eqref{eq:T1_chi} and \eqref{eq:bpar} yields the relaxation rate
    \begin{equation}
        \frac{1}{T_1} = \frac 83 \frac{\mu^2\omega^3}{\hbar c^3}b_\parallel^B(z,\omega) \coth\left(\frac{\hbar \omega}{2k_BT}\right).
        \label{eq:T1spin}
    \end{equation}
    Gaussian units are used throughout; in SI units the right-hand side should be multiplied by the factor $\mu_0/4\pi$. 

To get a sense of the behavior of the relaxation rates, it is useful to first consider the limit of a ``local medium", where the dielectric function $\epsilon(q,\omega) \to \epsilon(\omega)$, that is, a material where the response may be approximated as being independent of wave vector $q$. The local regime is sampled by qubits placed at a distance $z$ exceeding the intrinsic length scales of the material (e.g., the mean free path $\ell$). With a local dielectric function, the reflection coefficients for incident plane waves with frequency $\omega$ and wave vector $\mathbf Q$ reduce to the Fresnel expressions
    \begin{equation}
       r_s = \frac{v-\sqrt{\epsilon - u^2}}{v+ \sqrt{\epsilon - u^2}}, \quad r_p = \frac{\epsilon v-\sqrt{\epsilon - u^2}}{\epsilon v+ \sqrt{\epsilon - u^2}},
    \label{eq:fresnel}
    \end{equation}
where $u = c Q/\omega$ and $v=\sqrt{1-u^2}$, with  ${\rm Im}\sqrt{\epsilon - u^2} \geq 0$ and ${\rm Im} \, v\geq 0$; here $\epsilon = \epsilon(\omega)$.  As an example, suppose that the material is a good conductor, so that $|\epsilon| \gg 1$ ($\epsilon \approx 4\pi i \sigma/\omega$). Define the (local) skin depth by $\delta = c / \omega \sqrt{|\epsilon|}$ and consider the magnetic noise in the regime $z \ll \delta$. The $w^2r_s$ term dominates in Eq.~\eqref{eq:bpar} and expanding the integrand for $u \gg \sqrt{|\epsilon|}$, which is the dominant range when $z\ll\delta$, gives ${\rm Im}\,r_s\simeq {\rm Im}\,\epsilon/4u^2$ and hence
    \begin{equation}
        b_\parallel^B \approx \frac 34 \frac{{\rm Im}\epsilon(\omega)}{4} \int_0^\infty {\rm d} w ~ e^{-2 kz w} = \frac{3\pi}{8} \frac{c\sigma}{\omega^2} \frac{1}{z}.
    \label{eq:Bpar-local}
    \end{equation}
In the limit of a local medium, the magnetic noise is thus $\propto \sigma/z$, that is, it is directly proportional to the conductivity of the metal and falls off slowly with height. Equation~\eqref{eq:Bpar-local} requires $\ell\ll z\ll\delta$; for $z\gtrsim\delta$ the dominant partial waves are screened by the metal ($u^2\lesssim|\epsilon|$), ${\rm Im}\,r_s\simeq w\,{\rm Im}\,\epsilon/|\epsilon|^{3/2}$, and one finds instead the much steeper law $b^B_\parallel\simeq\tfrac94\,{\rm Im}\,\epsilon\,|\epsilon|^{-3/2}(kz)^{-4}$, i.e., $T_1\propto z^{4}$ (Appendix~\ref{app:half_space_greens_func}). The same $z^4$ law governs the superconducting state at distances where London screening dominates, $z\gtrsim\lambda/2$ with $\lambda$ the penetration depth, see Sec.~\ref{subsec:magnetic}. As we will see below, more generally the magnetic noise probes the nonlocal dissipative conductivity ${\rm Re}~\sigma(q,\omega)$ of the material. In particular, the temperature and height dependence of $T_1$ below $T_c$ can map out the properties of the nonlocal conductivity of a superconductor.

\subsection{Boundary conditions, nonlocal surface impedances, and reflection coefficients}

The general relationship between the current $j_i(\mathbf r)$ in a material and the electric field $E_i(\mathbf r)$ is nonlocal: $j_i(\mathbf r) = \int {\rm d}^3 \mathbf r' ~ \sigma_{ij}(\mathbf r, \mathbf r') E_j(\mathbf r')$. This relationship may be especially complicated near a boundary, where translation invariance is broken. The standard tractable model assumes that electrons reflect specularly from the surface (this is also known as the ``semiclassical infinite barrier" model \cite{Gerhardts1983,kliewer1968}). In this approximation, the half-space problem maps to the bulk problem and we make use of the bulk, translationally invariant response functions. The impedance is only weakly sensitive to the specularity assumption because the current is dominated by electrons moving nearly parallel to the surface \cite{reuter,fordweber}.

Assuming an isotropic dielectric medium, the bulk response separates into longitudinal ($l$) and transverse ($t$) parts:
    \begin{equation}
        \mathbf j(\mathbf q,\omega) = \sigma_l(q,\omega) \mathbf E_l + \sigma_t(q,\omega)\mathbf E_t, 
    \end{equation}
where $\epsilon_{l,t}(q,\omega) = 1 + 4\pi i \sigma_{l,t}(q,\omega)/\omega$.
Utilizing this decomposition, direct application of Maxwell's equation together with specular boundary conditions yields the reflection coefficients in terms of the dielectric function. In the case of $s$-polarized waves of frequency $\omega$ and lateral wave vector $\mathbf Q$, the result is
    \begin{equation}
        r_s(Q,\omega) = \frac{\zeta_s - \pi/v}{\zeta_s + \pi/v},
    \label{eq:rs}
    \end{equation}
where
    \begin{equation}
        \zeta_s(Q,\omega) = 2i \int_0^\infty \frac{{\rm d}y}{\epsilon_t(\omega\kappa/c,\omega)-\kappa^2},
    \label{eq:zetas}
    \end{equation}
with $\quad \kappa^2 = c^2Q^2/\omega^2+y^2$(so that $y$ is the dimensionless wave-vector component normal to the surface), and $v = \sqrt{1-c^2Q^2/\omega^2}$ (${\rm Im}\, v \geq 0$). For $p$-polarized waves, the result is
    \begin{equation}
        r_p(Q,\omega) = \frac{\pi v - \zeta_p}{\pi v + \zeta_p}
    \label{eq:rp}
    \end{equation}
where
    \begin{equation}
    \zeta_p(Q,\omega) = 2i \int_0^\infty \frac{{\rm d}y}{\kappa^2}\left[\frac{y^2}{\epsilon_t(\omega\kappa/c,\omega)-\kappa^2} + \frac{u^2}{\epsilon_l(\omega\kappa/c,\omega)}\right],
    \label{eq:zetap}
    \end{equation}
with $\kappa$ and $v$ defined as above. The quantities $\zeta_{s,p}/\pi$ are the nonlocal surface impedances of the medium. In the local limit, $\epsilon_{t,l}\to\epsilon(\omega)$, the integrals are elementary, $\zeta_s\to\pi/\sqrt{\epsilon-u^2}$ and $\zeta_p\to\pi\sqrt{\epsilon-u^2}/\epsilon$, and Eqs.~\eqref{eq:rs} and \eqref{eq:rp} reduce to the Fresnel coefficients \eqref{eq:fresnel}. The details of the derivation of Eqs.~\eqref{eq:rs} and \eqref{eq:rp} can be found in Appendix~\ref{app:nonlocal_impedances}. 

\subsection{Normal-state nonlocal response}
\label{subsec:normal_state_response}

Before discussing the superconducting state, in this subsection we recall the normal-metal response functions. In the range of wave vectors $q \ll k_F$ of interest here (corresponding to qubit probe distance $z \gg 1/k_F$), the response to an electric field $\mathbf E$ can be obtained from the Boltzmann equation in the relaxation-time approximation,
    \begin{equation}
        \frac{\partial f}{\partial t} + \mathbf v \cdot \frac{\partial f}{\partial \mathbf r} +e \mathbf E \cdot \frac{\partial f}{\partial \mathbf p} = - \frac{f-f_0}{\tau},
    \end{equation}
where $f_0$ is the equilibrium distribution function and we have set $\dot{\mathbf p} = e\mathbf E$.

Consider first the response to a transverse field $\mathbf E = E\hat x  e^{i(qz - \omega t)}$ ($\mathbf E \perp \mathbf q$). Near equilibrium, we set $f=f_0+\delta f$ and obtain
\begin{equation}
(-i\omega + i q\vF n_z+1/\tau)\,\delta f
=eE\,n_x \vF\,\big(-\partial_\ee f_0\big),
\end{equation}
where $\mathbf n$ is the direction on the Fermi surface and $v_F$ is the Fermi velocity.  The current is
$j_x=2e\nuF\vF\langle n_x\,\delta f\rangle_n$ (the average being taken over the Fermi surface), where $\nu_F$ is the density of states on the Fermi surface (per spin) and the factor 2 is from a sum over spins. Setting $\sigma_t = j_x / E_x$
and defining the kernel $\hat Q_N\equiv- i \omega\sigma_t / (ne^2/m)$ we find
\begin{equation}
\hat Q_N(q, \omega) =
\frac{3\omega}{4\vF q}\,\mathcal J\!\Big(\frac{\omega+\ii/\tau}{\vF
q}\Big),
\label{eq:QN}
\end{equation}
where
    \begin{equation}
        \mathcal J(x) \equiv (1-x^{2})\ln\frac{x+1}{x-1}+2x,
    \label{eq:Jdef}
    \end{equation}
with the principal branch of the logarithm (${\rm Im}\,x>0$).
The transverse dielectric function is
    \begin{equation}
        \eps_t(q,\omega)=1-\omega_p^2\hat Q_N/\omega^2,
    \end{equation}
where $\omega_p^2 = 4\pi ne^2/m$ is the plasma frequency.

Note the limiting behaviors: When $q\to0$, $\mathcal J\to4/3x$ and we recover the Drude result
$\hat Q_N=\omega/(\omega+\ii/\tau)$, $\eps_t=1-\omega_p^2/\omega(\omega+
\ii/\tau)$. For $\vF q\gg\omega,1/\tau$, $\mathcal J\to -\ii\pi$ and
$\im\eps_t\simeq (3\pi/4) \omega_p^2/\omega v_F q\,>0$,
independent of $\tau$: the surface responds as a collisionless electron gas
(Landau damping).

In the longitudinal channel, the density response requires particle conservation to be respected in the
presence of collisions. This can be achieved using Mermin's  construction
\cite{mermin}, which combines the collisionless semiclassical (Fermi-surface) dielectric function $\eps^{0}_l(q,\omega)=1+({q_{\rm TF}^{2}}/{q^{2}})[1-\tfrac{x}{2}\ln\frac{x+1}{x-1}]$, with $x=\omega/q\vF$ and $q_{\rm TF}^{2}=3\omega_p^2/\vF^2$, evaluated at the complex frequency $\omega+\ii/\tau$, with a relaxation-time correction that conserves the local particle number. The result can be written compactly as
\begin{equation}
\eps_l(q,\omega)=1+\frac{q_{\rm TF}^{2}}{q^{2}}\,
\frac{1+\dfrac{\omega+\ii/\tau}{2q\vF}
\ln\dfrac{\omega-q\vF+\ii/\tau}{\omega+q\vF+\ii/\tau}}
{1+\dfrac{\ii/\tau}{2q\vF}
\ln\dfrac{\omega-q\vF+\ii/\tau}{\omega+q\vF+\ii/\tau}}\;,
\label{eq:mermin2}
\end{equation}
whose $q\to0$ limit is again the Drude result. At the small frequencies of interest the real part is dominated by
Thomas--Fermi screening, $\re ~ \eps_l\simeq1+q_{\rm TF}^{2}/q^{2}$, while
$\im~ \eps_l\propto\omega$ describes Landau-damped charge fluctuations.

We stress that the transverse and longitudinal
functions differ substantially in the nonlocal regime (even their
$q$-asymptotics differ), and they enter different places:
$\eps_t$ controls $\zeta_s$ (magnetic noise) and the $y^{2}$ term of
$\zeta_p$; $\eps_l$ controls the electrostatic term of $\zeta_p$ (electric
noise).

\section{Results}
\label{sec:results}

We now turn to EWJN above a bulk superconductor. We consider the case of an $s$-wave superconductor with both point-like non-magnetic and paramagnetic impurities; the clean system is discussed as a limiting case. The results in this section require the nonlocal optical conductivity (equivalently, the transverse current-current response kernel, $Q(q,\omega) = -i\omega \sigma_t(q,\omega)$) below $T_c$, which we calculate within BCS mean-field theory. Within this context, the results below are derived  for arbitrary $\omega, T, q \ll k_F$, and impurity concentrations. The structure of the final result agrees with the kernel of
Ref.~\cite{kharitonov}; our derivation is independent, and in one place [Eq.~\eqref{eq:hcorrect}] we obtain a result that corrects Ref.~\cite{kharitonov} beyond the Born limit.

The starting mean-field model is
\begin{equation}
\begin{aligned}
\hat H &=\!\int_{z<0}\!\dd^{3}r\,\Big\{\hat\psi^{\dagger}_{\sigma}
\Big[E\big(\hat{\bm p}-\tfrac{e}{c}\bm A\big)-\ee_F
+\sum_a u\,\delta(\bm r-\bm r_a)\Big]\hat\psi_{\sigma}
\\
&\qquad +\sum_b J\,\bm s_b\!\cdot\!\big(\hat\psi^{\dagger}\bm\sigma\hat\psi\big)
\delta(\bm r-\bm r_b)
+\Delta\big[\hat\psi_{\uparrow}\hat\psi_{\downarrow}+\text{h.c.}\big]\Big\},
\end{aligned}
\end{equation}
with isotropic spectrum $E(p)$, unpolarized classical impurity spins $\bm s_b$ ($|\bm s_b|=1$), and the order parameter $\Delta$ determined self-consistently.  Disorder is treated in the standard noncrossing approximation \cite{agd}. Magnetic impurities are treated to all orders in $J$ for a single impurity (self-consistent $T$-matrix), i.e.\ in the Shiba approximation \cite{shiba1968,*shiba1973}.  Optimal-fluctuation ``tails'' of the density of states \cite{lamacraft} are beyond this approximation and are not included.

\subsection{Nambu propagator and non-magnetic disorder}

In the Nambu basis $\Psi_p=(\psi_{p\uparrow},\psi^{\dagger}_{-p\downarrow})$
the disorder-averaged retarded Green function has the form
\begin{equation}
\mathsf G^{R}(\ee,\bm p)
=\frac{\tilde\ee\,\tau_0+\xi_p\,\tau_3+\tilde\Delta\,\tau_1}
{\tilde\ee^{2}-\xi_p^{2}-\tilde\Delta^{2}},
\quad \xi_p=E(p)-\ee_F ,
\label{eq:GNambu}
\end{equation}
with renormalized quasiparticle energy $\tilde\ee(\ee)$, and gap function $\tilde\Delta(\ee)$. We define
\begin{equation}
D(\ee)=\sqrt{\tilde\ee^{2}-\tilde\Delta^{2}}\quad(\im D\ge0),
\end{equation}
and $g = \tilde\ee / D$, $f=\tilde\Delta / D$, so $g^{2}-f^{2}=1$. The local (momentum-integrated) propagator needed below is
\begin{equation}
\mathfrak g_0(\ee)\equiv\nuF\!\int\!\dd\xi\;\mathsf G^{R}(\ee,\bm p)
=-\ii\pi\nuF\,\big(g\,\tau_0+f\,\tau_1\big),
\label{eq:g0local}
\end{equation}
where we used $\int\dd\xi\,(D^{2}-\xi^{2})^{-1}=-\ii\pi/D$ (the pole is in the upper half
plane at $\xi=D$) and dropped the particle--hole--asymmetric $\tau_3$ piece.
The density of states is $\nu(\ee)=\nuF\,\re \, g(\ee)$.

In the case of non-magnetic impurities of density $n$ and potential $u$ (the corresponding vertex being $u\tau_3$), the self-energy in the Born approximation is
\begin{equation}
\Sigma_{n}=n u^{2}\,\tau_3\,\mathfrak g_0\,\tau_3
=-\frac{\ii}{2\tau}\big(g\,\tau_0-f\,\tau_1\big),
\quad \frac1\tau=2\pi\nuF n u^{2},
\end{equation}
so that $\tilde\ee=\ee+\ii g/2\tau$ and
$\tilde\Delta=\Delta+\ii f/2\tau$. Both functions are renormalized
with the same sign and the ratio $\tilde\ee/\tilde\Delta$ is unchanged by disorder. The density of states is similarly unaffected, which is a manifestation of Anderson's theorem
\cite{anderson}.  The only effect is
\begin{equation}
D\;\longrightarrow\;D_s+\frac{\ii}{2\tau},
\label{eq:Dshift}
\end{equation}
where $D_s$ contains the magnetic renormalization to be calculated next.

\subsection{Magnetic impurities: $T$-matrix to all orders}

Turning to magnetic impurities of density $n_s$ and exchange $J$, we first observe that, for a classical spin, the exchange potential is diagonal when the electron spin
is quantized along $\bm s_b$. In the corresponding Nambu block it reads
$J\tau_0$ (particle and Bogoliubov-hole components shift the same way and it is this sign structure, which is opposite to $u\tau_3$, that makes magnetic scattering pair-breaking).  The single-impurity $T$-matrix is the full Born series
\begin{equation}
\mathsf T=J\tau_0+J\tau_0\,\mathfrak g_0\,\mathsf T.
\end{equation}
Solving for $\mathsf T$, we obtain
\begin{equation}
\mathsf T(J)=J\,\frac{(1+\ii\alpha g)\tau_0-\ii\alpha f\,\tau_1}
{1+2\ii\alpha g-\alpha^{2}},
\quad \alpha\equiv\pi\nuF J.
\label{eq:Tmatrix}
\end{equation}
The poles of \eqref{eq:Tmatrix} inside the gap reproduce the
Yu-Shiba-Rusinov bound state at
\begin{equation}
\ee_0=\gamma\Delta,
\qquad
\gamma\equiv\frac{1-\alpha^{2}}{1+\alpha^{2}}.
\label{eq:shibastate}
\end{equation}
Averaging over the (unpolarized, isotropic) spin
directions is equivalent to averaging $\mathsf T(J)$ and $\mathsf T(-J)$; in the full Nambu$\otimes$spin space this follows by writing $J\tau_0\mathfrak g_0=M\otimes W$ with $M=\bm s\cdot\bm\sigma$, $M^2=1$, $\langle M\rangle_{\bm s}=0$, and $W=-\ii\alpha(g\tau_0+f\tau_1)$, so that the spin-flip components drop out of the average and
\begin{align}
\big\langle \mathsf T\big\rangle_{\bm s}
&=J\,W\,(1-W^{2})^{-1} \nonumber\\
&=-\ii\alpha J\,
\frac{g(1+\alpha^{2})\,\tau_0+f(1-\alpha^{2})\,\tau_1}
{(1-\alpha^{2})^{2}+4\alpha^{2}g^{2}} .
\end{align}
With $(1-\alpha^{2})^{2}+4\alpha^{2}g^{2}
=(1+\alpha^{2})^{2}\big[\gamma^{2}+(1-\gamma^{2})g^{2}\big]$ the
self-energy $\Sigma_s=n_s\langle\mathsf T\rangle_{\bm s}$ gives the renormalized quantities
\begin{equation}
\begin{aligned}
\tilde\ee&=\ee+\frac{\ii}{2\tau_s}\,
\frac{(1+\alpha^{2})\,g}{\gamma^{2}+(1-\gamma^{2})g^{2}}, \\
\tilde\Delta&=\Delta-\frac{\ii}{2\tau_s}\,
\frac{(1+\alpha^{2})\,\gamma f}{\gamma^{2}+(1-\gamma^{2})g^{2}},
\label{eq:selfenergies}
\end{aligned}
\end{equation}
where
\begin{equation}
\frac{1}{\tau_s}=\frac{2\pi\nuF n_s J^{2}}{(1+\alpha^{2})^{2}}
=\frac{n_s}{2\pi\nuF}\,(1-\gamma^{2}) .
\end{equation}
Note the sign structure of \eqref{eq:selfenergies}: the $\tau_0$ and
$\tau_1$ channels are renormalized with opposite signs (pair
breaking), and beyond the Born limit with different magnitudes
(the extra factor $\gamma$ in the $\tilde\Delta$ channel).

The standard Shiba equation \cite{shiba1968,*shiba1973,kharitonov} is obtained by taking the ratio $v=\tilde\ee/\tilde\Delta$ (so that $g=v/\sqrt{v^{2}-1}$,
$f=1/\sqrt{v^{2}-1}$) and using Eq.~\eqref{eq:selfenergies} together with the identity $(1+\alpha^2)(1+\gamma)=2$:
\begin{equation}
v\,\Delta=\ee+\frac{1}{\tau_s}\,
\frac{v\,\sqrt{1-v^{2}}}{\gamma^{2}-v^{2}}.
\label{eq:shibaeq}
\end{equation}
Here $D_s$ denotes $\sqrt{\tilde\ee^{2}-\tilde\Delta^{2}}$ evaluated with the magnetic self-energies \eqref{eq:selfenergies} alone; the non-magnetic shift is added afterwards according to Eq.~\eqref{eq:Dshift}. Eliminating the self-energies gives the closed form
\begin{equation}
D_s(\ee)=\frac{\ee}{g(\ee)}
+\frac{\ii}{\tau_s}\,\frac{v^{2}-1}{(1+\gamma)\,(v^{2}-\gamma^{2})}.
\label{eq:hcorrect}
\end{equation}
In the Born (Abrikosov-Gor'kov) limit $\gamma\to1$ this reduces to
$D_s=\ee/g+\ii/2\tau_s$, which coincides with the relation
$D_s=\ee/2g+\Delta/2f$ quoted in Ref.~\cite{kharitonov},
Eqs.~(4.13)--(4.14).  Beyond the Born limit, however, Eq.~\eqref{eq:hcorrect} corrects the results of Ref.~\cite{kharitonov}: solving the coupled self-consistency equations \eqref{eq:selfenergies} numerically, we find that the two forms agree at $\gamma=1$ but differ at the tens-of-percent level at $\gamma=0$. Since $D_s$ (together with
the shift \eqref{eq:Dshift}) sets the pole positions of the propagator and hence the momentum structure of the response, the correction matters for stronger exchange coupling, $\alpha=\pi\nuF J\sim1$.  Everywhere below,
$h(\ee)\equiv D_s(\ee)$ is computed from \eqref{eq:hcorrect} (or,
equivalently, directly from the converged
$\sqrt{\tilde\ee^{2}-\tilde\Delta^{2}}$). We note that both terms of \eqref{eq:hcorrect} are purely imaginary wherever the density of states vanishes ($g$ imaginary, i.e.\ $v$ real with $|v|<1$); this property guarantees the vanishing of dissipation in the gapped region (Appendix~\ref{app:checks}).

\begin{figure}[t]
\centering
\includegraphics[width=\columnwidth]{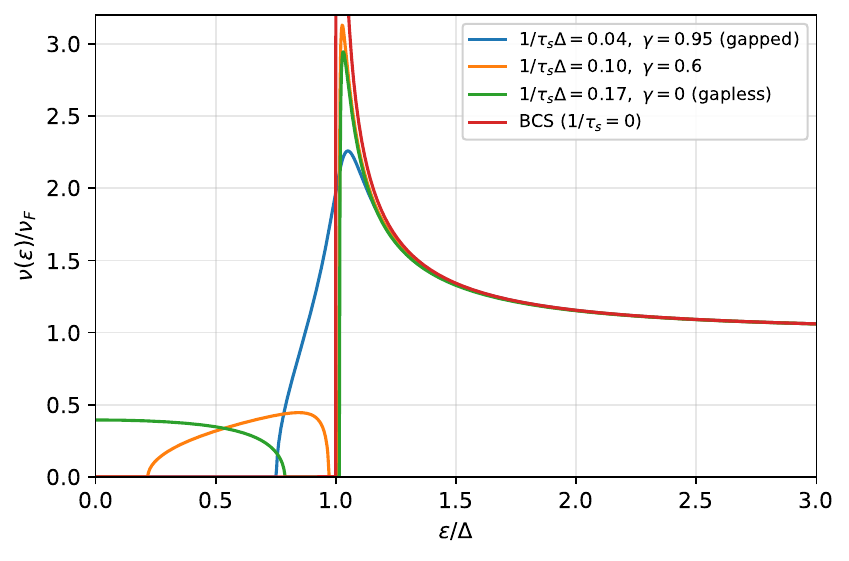}
\caption{Quasiparticle density of states from the Shiba theory,
Eqs.~\eqref{eq:selfenergies}--\eqref{eq:shibaeq}, for representative
parameters (cf.\ Ref.~\cite{kharitonov}, Fig.~1). Energies are measured in units of the self-consistent gap $\Delta$.}
\label{fig:dos}
\end{figure}

Solving \eqref{eq:shibaeq} and taking $\nu(\ee)=\nuF\re \, g$ produces the familiar phenomenology, shown in Fig.~\ref{fig:dos}: for weak coupling and low impurity
concentration a hard gap survives with broadened edges and an impurity band
around $\ee_0=\gamma\Delta$; for stronger coupling and/or scattering the gap
closes, with $\nu(0)>0$ while
$\Delta\neq0$, leading to gapless superconductivity.

\subsection{Self-consistent gap and $T_c$ suppression}

\begin{figure}[t!]
\centering
\includegraphics[width=\columnwidth]{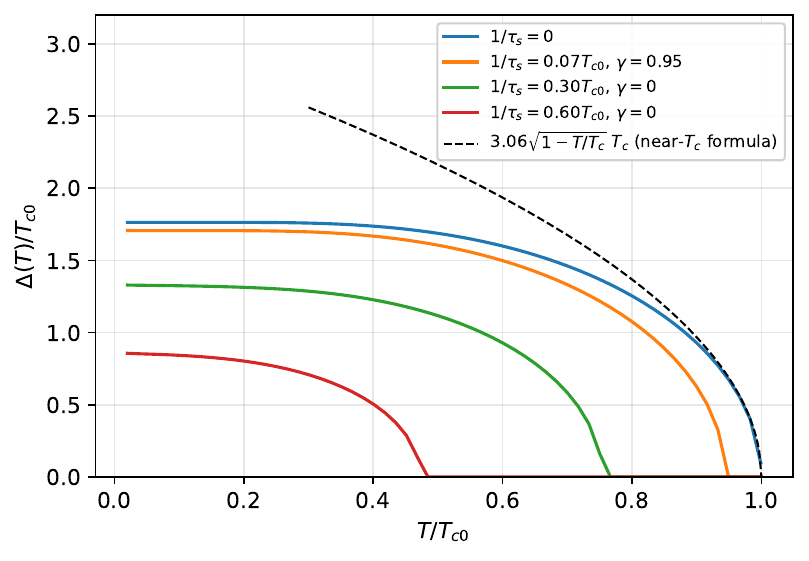}
\caption{Self-consistent gap with magnetic pair breaking, from
Eqs.~\eqref{eq:shibamatsu}--\eqref{eq:gapeq}.  The dashed curve is the
near-$T_c$ formula, shown to emphasize its limited domain of validity.}
\label{fig:gap}
\end{figure}

On the Matsubara axis ($\ee\to\ii\omega_n$, $\omega_n=2\pi T(n+\tfrac12)$,
$v\to\ii u_n$) the Shiba equation becomes real,
\begin{equation}
u_n\Delta=\omega_n+\frac{1}{\tau_s}\,
\frac{u_n\sqrt{1+u_n^{2}}}{\gamma^{2}+u_n^{2}},
\label{eq:shibamatsu}
\end{equation}
and the linearized-coupling gap equation, with the coupling constant
eliminated in favor of the clean transition temperature $T_{c0}$, reads
\begin{equation}
\ln\frac{T_{c0}}{T}=2\pi T\sum_{n\ge0}
\left[\frac{1}{\omega_n}-\frac{1}{\Delta\sqrt{1+u_n^{2}}}\right].
\label{eq:gapeq}
\end{equation}
In the clean limit, $u_n=\omega_n/\Delta$ recovers BCS,
$\Delta_0 \equiv \Delta(T = 0)\approx 1.764\,T_{c0}$.  Near
$T_c$, $u_n\Delta\to\omega_n+1/\tau_s$ independently of $\gamma$, and
\eqref{eq:gapeq} reduces to the Abrikosov-Gor'kov formula
\begin{equation}
\ln\frac{T_{c0}}{T_{c}}
=\psi\!\Big(\frac12+\frac{1}{2\pi\tau_sT_{c}}\Big)-\psi\!\Big(\frac12\Big),
\label{eq:AGTc}
\end{equation}
with complete suppression at $1/\tau_s^{*}=\Delta_0/2\approx 0.882\,T_{c0}$
\cite{ag}. For the pair-breaking rates used below, Eq.~\eqref{eq:AGTc} gives $T_c/T_{c0} \approx 0.75$ for $1/\tau_s=0.30\,T_{c0}$ and $T_c/T_{c0}\approx 0.47$ for $1/\tau_s=0.60\,T_{c0}$. Figure~\ref{fig:gap} shows $\Delta(T)$ for the parameter sets used later.  The frequently quoted near-$T_c$ formula $\Delta\approx 3.06\sqrt{T_c(T_c-T)}$ is also shown in the figure, which is seen to be quantitatively inaccurate for $T\lesssim 0.9\,T_c$. Since $\Delta$ enters exponentially at low $T$, we will use the full numerical solution of Eq.~\eqref{eq:gapeq}.

\subsection{Response functions}

Turning now to the transverse conductivity, the transverse kernel is defined by $\bm j(\bm q,\omega)=-\tfrac1c Q(q,\omega)\,
\bm A(\bm q,  \omega)$ for transverse $\bm A$ ($\bm A\cdot\bm q=0$), and
$Q=-\ii\omega\sigma_t$. The corresponding transverse dielectric function is given by
\begin{equation}
\eps_t(q,\omega)=1-\frac{4\pi Q(q,\omega)}{\omega^{2}}
=1-\frac{\omega_p^{2}}{\omega^{2}}\hat Q(q,\omega),
\label{eq:epsfromQ}
\end{equation}
where  $\hat Q\equiv  Q / Q_0$ and $Q_0 = ne^2/m = 2(e\vF)^{2}\nuF/3$ (cf. Sec.~\ref{subsec:normal_state_response}). The limit
$Q_0=Q(0,0)|_{T=0,\rm clean}$ gives the London penetration depth $\lambda_{L0}^{-2}=4\pi Q_0/c^{2}=\omega_p^{2}/c^{2}$.

Standard analytic continuation of the current-current bubble to real
frequencies (see, e.g., \cite{agd}) expresses $Q$ through the retarded and
advanced disorder-averaged propagators.  For point disorder and a transverse
vertex the ladder vertex corrections vanish identically---each rung carries
$\langle \bm n\rangle$-odd factors---so the bubble is the product of averaged
Green functions:
\begin{equation}
\begin{aligned}
\hat Q(k,\omega) = -\frac{3\ii}{2}\!&\int_{-\infty}^{\infty}\!\!\dd\ee
\left\{(\thf_+-\thf_-)\,\Pi^{RA}(\ee) \right. \\
&\qquad \left. +\thf_-\,\Pi^{RR}(\ee)-\thf_+\,[\Pi^{RR}(\ee)]^{*}\right\}.
\label{eq:Qmaster}
\end{aligned}
\end{equation}
Here $\thf_\pm=\tanh[(\ee\pm\omega/2)/2T]$ and
$\Pi^{ab}$ are the $\xi$-integrated, angle-averaged products of two
propagators at energies $\ee_\pm=\ee\pm\omega/2$ and momenta
$\bm p\pm\bm k/2$.

With $\mathsf G^{R}$ of Eq.~\eqref{eq:GNambu} written as
$G^{R}=(\tilde\ee+\xi)/(\tilde\ee^{2}-\xi^{2}-\tilde\Delta^{2})$,
$F^{R}=\tilde\Delta/(\tilde\ee^{2}-\xi^{2}-\tilde\Delta^{2})$, and
$\xi_{\bm p\pm\bm k/2}\simeq\xi\pm b$, $b=\tfrac12\vF\,\bm n\cdot\bm k$, the
required integrals follow from residues.  Denoting
$D_\pm=D(\ee_\pm)=h(\ee_\pm)+\ii /2\tau$ [Eqs.~\eqref{eq:Dshift},
\eqref{eq:hcorrect}]:
\begin{subequations}
\begin{align}
\int\frac{\dd\xi}{2\pi}\Big[G^{R}_+G^{A}_-+F^{R}_+F^{A}_-\Big]
&\doteq\frac{\ii}{2}\,
\frac{1+g_+g_-^{*}+f_+f_-^{*}}{D_+-D_-^{*}-2b} \nonumber
\\
&\equiv\;C_{RA}\,\frac{\ii}{D_+-D_-^{*}-2b},
\label{eq:xiRA}\\
\int\frac{\dd\xi}{2\pi}\Big[G^{R}_+G^{R}_-+F^{R}_+F^{R}_-\Big]
&\doteq\frac{\ii}{2}\,
\frac{1-g_+g_--f_+f_-}{D_++D_--2b} \nonumber \\
&\equiv\;C_{RR}\,\frac{\ii}{D_++D_--2b}.
\label{eq:xiRR}
\end{align}
\end{subequations}
The symbol $\doteq$ means equality after symmetrization
$b\to-b$, which is permitted because the angular average below is even in $b$. The combinations
\begin{equation}
C_{RA}=\tfrac12\big(1+g_+g_-^{*}+f_+f_-^{*}\big),\quad
C_{RR}=\tfrac12\big(1-g_+g_--f_+f_-\big)
\label{eq:cohfactors}
\end{equation}
are the BCS coherence factors of the transverse current vertex.

The transverse projection gives $\langle n_\alpha^{2}(\cdots)\rangle_{\bm n}
=\tfrac14\int_{-1}^{1}\dd x\,(1-x^{2})(\cdots)$. Evaluating the resulting elementary integral, we obtain
\begin{align}
\mathcal A(l_0)&\equiv\Big\langle n_\alpha^{2}\,
\frac{\ii}{\vF k\,(l_0-n_z)}\Big\rangle_{\bm n} \nonumber \\
&=\frac{-\ii}{4\vF k}\Big[(1-l_0^{2})\ln\frac{l_0-1}{l_0+1}-2l_0\Big],
\label{eq:Afun}
\end{align}
which is analytic for $\im \, l_0>0$, and has the limiting behaviors $\mathcal A\to\ii/(3\vF k\,l_0)$ for
$|l_0|\to\infty$ and $\mathcal A\to\pi/(4\vF k)$ for $l_0\to0$.

Putting everything together, we arrive at the central result
\begin{align}
\hat Q(k,\omega)=-\frac{3\ii}{2}\int_{-\infty}^{\infty}\dd\ee\,
\left\{(\thf_+-\thf_-)\,C_{RA}\,\mathcal A\big(l^{RA}_0\big) \right. \nonumber \\  
\left.+\thf_-\,C_{RR}\,\mathcal A\big(l^{RR}_0\big)
-\thf_+\,\big[C_{RR}\,\mathcal A\big(l^{RR}_0\big)\big]^{*}\right\},
\label{eq:kernelfinal}
\end{align}
where
\begin{subequations}
\label{eq:l0def}
\begin{align}
&l^{RA}_0=\frac{h(\ee_+)-h^{*}(\ee_-)+\ii/\tau}{\vF k}, \\
&l^{RR}_0=\frac{h(\ee_+)+h(\ee_-)+\ii/\tau}{\vF k}.
\end{align}
\end{subequations}
Equation \eqref{eq:kernelfinal} together with
Eqs.~\eqref{eq:selfenergies}--\eqref{eq:hcorrect} and
Eq.~\eqref{eq:gapeq} is the complete microscopic input to the noise
calculation.  It has the same structure as Eqs.~(4.1), (4.15)--(4.17) of
Ref.~\cite{kharitonov}, with $h$ replaced by the corrected
\eqref{eq:hcorrect}. In Appendix \ref{app:checks}, we verify that Eq.~\eqref{eq:kernelfinal} reduces to the established forms of the transverse current response function in the normal-state (Drude-Boltzmann), London, Pippard, and Mattis-Bardeen limits, and we derive the residual dissipation in the gapless regime, which is the source of the noise floor discussed below.

In the next subsections, we present our results for the magnetic noise, which are obtained from Eq.~\eqref{eq:kernelfinal} by direct numerical integration. We also comment briefly on charge qubit relaxation due to electric noise. Details of the numerical procedure are summarized in Appendix \ref{app:numerics}.

\subsection{Magnetic noise and spin-qubit $T_1$}
\label{subsec:magnetic}

\begin{table}[b]
\centering
\begin{tabular}{lll}
\toprule
quantity & symbol & range considered \\
\midrule
temperature & $T$ & $0.05$--$1\,T_c$ ($T_c=9.26\,$K)\\
qubit frequency & $\omega/2\pi$ & GHz  (default 2.88 GHz)\\
qubit height & $z$ & $2$--$100$~nm\\
elastic scattering & $\tau T_c$ & $0.02$--$20$\\
magnetic scattering & $1/\tau_s$ & $0$--$0.9\,T_{c0}$\\
exchange strength & $\gamma=\frac{1-(\pi\nuF J)^2}{1+(\pi\nuF J)^2}$ & $0$--$1$\\
coherence length & $\xi_0=\vF/\pi\Delta_0$ & $38$~nm\\
London depth & $\lambda_{L0}=c/\omega_p$ & $39$~nm\\
mean free path & $\ell=\vF\tau$ & $4.2$~nm--$4.2\,\mu$m\\
residual resistivity & $\rho=4\pi/\omega_p^2\tau$ & $0.01$--$12\,\mu\Omega\,$cm\\
\bottomrule
\end{tabular}
\caption{Parameters.  Energies and rates are quoted in units where
$\hbar=\kB=1$; e.g.\ $T_c\equiv \kB T_c/\hbar=1.213\times10^{12}\,$s$^{-1}$, $\Delta_0=1.764\,T_c=2.140\times10^{12}\,$s$^{-1}$ and $k_F \approx 1.2 \times 10^{10} $m$^{-1}$ for Nb. The niobium input parameters are the measured $T_c=9.26$~K \cite{finnemore1966}, London penetration depth $\lambda_{L0}=39$~nm, and coherence length $\xi_0=38$~nm \cite{finnemore1966,tinkham}, which fix the two quantities that enter the kernel \eqref{eq:kernelfinal} and the surface impedances, $\omega_p=c/\lambda_{L0}=7.7\times10^{15}$~s$^{-1}$ and $\vF=\pi\Delta_0\xi_0/\hbar=2.6\times10^{5}$~m/s (an effective Fermi velocity of the multiband Fermi surface of Nb). The value of $k_F$ enters only the validity condition $q\ll k_F$. The range of $\tau T_c$ corresponds to mean free paths from $4$~nm to $4\,\mu$m, i.e., from strongly disordered sputtered films ($\tau T_c=0.02$, $\rho\approx12\,\mu\Omega\,$cm) through typical epitaxial films ($\tau T_c=0.2$, $\ell=42$~nm, $\rho\approx1.2\,\mu\Omega\,$cm) to bulk single crystals ($\tau T_c=20$).}
\label{tab:params}
\end{table}

We begin with the magnetic noise below $T_c$ and the corresponding spin-qubit relaxation rate $1/T_1$. For concreteness, we will use material parameters appropriate to niobium, which we take as a representative conventional $s$-wave superconductor, when presenting our results. These materials parameters are summarized in Table \ref{tab:params}. For the qubit, we will use the typical NV qubit frequency $\omega/2\pi=2.88$~GHz, and spin-1/2 matrix element $\mu=\mu_B$ with spin quantization along the surface normal, so that Eq.~\eqref{eq:T1spin} applies. (For the spin-1 NV center the transverse matrix elements are larger by $\sqrt2$, so that all rates quoted below should be doubled; the temperature and distance dependences are unaffected. The NV axis is inclined with respect to the normal of a (100) diamond surface, so that the measured rate is a weighted combination of $\chi_\parallel$ and $\chi_\perp$; since both components have the same temperature dependence in the near field, none of the conclusions below is affected.) All results are for a half-space; they apply to a film of thickness $d$ provided $d$ exceeds the larger of the penetration depth and the skin depth at the relevant wave vectors, $d\gtrsim100$~nm for Nb, while for thinner films the reflection coefficients \eqref{eq:rs} and \eqref{eq:rp} must be replaced by their thin-film (stack) generalizations.

\begin{figure*}[t]
\centering
\includegraphics[width=\textwidth]{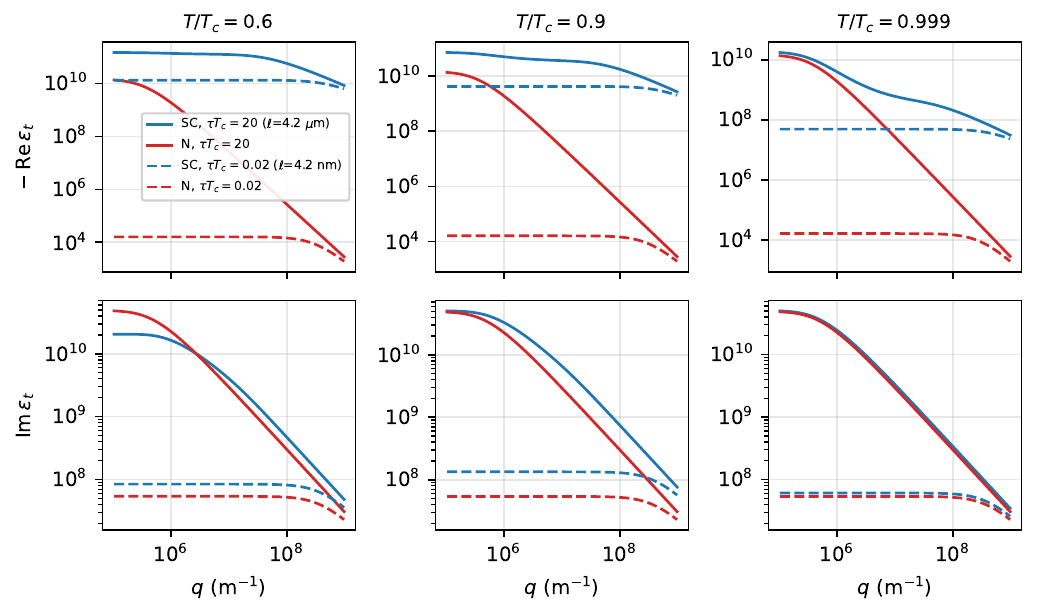}
\caption{Transverse dielectric function of Nb at $2.88$~GHz,
for the superconducting (blue) vs.\ normal (red) states. Solid lines are for clean samples ($\tau T_c=20$, $\ell=4.2\,\mu$m, $1/\ell=2.4 \times 10^5~{\rm m}^{-1}$) and
dashed lines are for dirty samples ($\tau T_c=0.02$, $\ell=4.2$~nm, $1/\ell=2.4 \times 10^8~{\rm m}^{-1}$). $1/\xi_0 = 2.6 \times 10^{7}~{\rm m}^{-1}$. Three temperatures are shown. The plotted range $10^{5}\,{\rm m}^{-1}\le q\le10^{9}\,{\rm m}^{-1}$ corresponds to the dominant momenta $q\sim1/2z$ for qubit heights between $\sim1$~nm and several micrometers, i.e., the experimentally relevant range; the upper end is still an order of magnitude below $k_F$, where the quasiclassical kernel applies.}
\label{fig:eps}
\end{figure*}

Figure~\ref{fig:eps} shows the transverse dielectric function $\epsilon_t$ entering $\zeta_s$, in both the superconducting and normal states for purposes of comparison.  The main features are readily understood from the limiting behaviors. At the relatively small $\omega$ considered here there is a large negative real part in the superconducting state at small $q$ because of the superfluid/London screening, $-\re ~ \eps_t\simeq c^{2}/\lambda^{2}\omega^{2}$. As to the $q$ dependence, there is a $1/q$ decay for $ q\gtrsim 1/\xi_{\rm eff}$, which is the nonlocal Pippard correction; here $\xi_{\rm eff}$ is Pippard's effective coherence length, $\xi_{\rm eff}^{-1} = \xi_0^{-1} + \ell^{-1}$ \cite{pippard1953,tinkham}, which interpolates between the clean-limit BCS coherence length $\xi_0=\vF/\pi\Delta_0$ and the mean free path $\ell$ in the dirty limit. The dissipative part $\im ~ \eps_t$ is suppressed relative to the normal state by the quasiparticle population and coherence factors, with the suppression strongest at low $T$, all consistent with a simple two-fluid picture.
Finally, as $T\to T_c^-$, it can be seen that the superconducting curves smoothly approach the normal state results.  The mean free path and coherence length are visible as the crossover scales $q\sim1/\ell$ and $q\sim1/\xi_0$ in the clean case, while in the dirty case ($\tau T_c=0.02$, $\ell\ll\xi_0$) the single scale $1/\ell$ controls the crossover.

\begin{figure}[t!]
\centering
\includegraphics[width=\columnwidth]{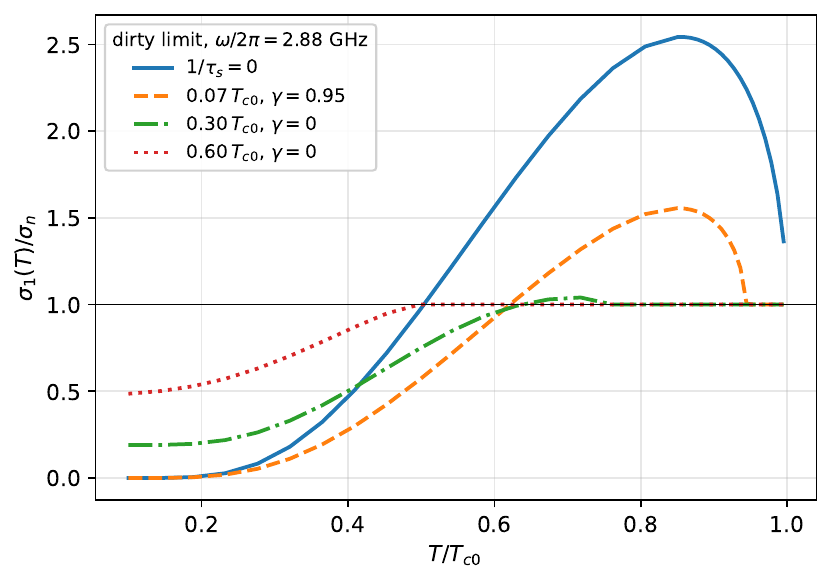}
\caption{Dirty-limit $\sigma_1(T)/\sigma_n$ at 2.88 GHz as computed from the kernel
\eqref{eq:kernelfinal}, for $\tau T_c=0.02$ ($1/\tau\approx28\,\Delta_0$); in this $k\to0$, $1/\tau\gg\Delta$ limit the ratio $\sigma_1/\sigma_n$ is independent of $\tau$ (Mattis--Bardeen) and of the material parameters other than $\Delta(T)$. As magnetic scattering is turned on the coherence peak disappears and the system becomes gapless. In the gapless regime the residual dissipation satisfies
$\sigma_1(0)/\sigma_n=[\nu(0)/\nuF]^{2}$.}
\label{fig:sigma1}
\end{figure}

In Figure~\ref{fig:sigma1} we report the local ($k \to 0$) dissipative conductivity $\sigma_1(T)/\sigma_n$ in the dirty limit, with and without magnetic impurities.  We highlight the features that are important for the noise: First, there is a prominent Hebel-Slichter-type coherence peak, with a maximum $\sigma_1/\sigma_n\simeq2.5$ at
$T\simeq0.85T_c$ for $1/\tau_s=0$, in agreement with the
Mattis--Bardeen result (the peak is described by the asymptote given in Appendix~\ref{app:checks}, Eq.~\eqref{eq:HSasympt}). The exponential decrease of the dissipation at
low $T$ is also visible. Finally, there are two notable features in the presence of magnetic impurities: (i) the rapid destruction of the
coherence peak (pair breaking broadens the DOS singularity that drives it) and (ii) saturation of $\sigma_1$ at the residual value
$[\nu(0)/\nuF]^{2}\sigma_n$ in the gapless regime (cf.
Eq.~\eqref{eq:gaplessfloor} at $k\to0$); numerically, $\sigma_1(T\to0)/\sigma_n \approx 0.19$ and $0.49$ for the two gapless curves, compared with $[\nu(0)/\nuF]^2 \approx 0.20$ and $0.50$.

\begin{figure}[t!]
\centering
\includegraphics[width=\columnwidth]{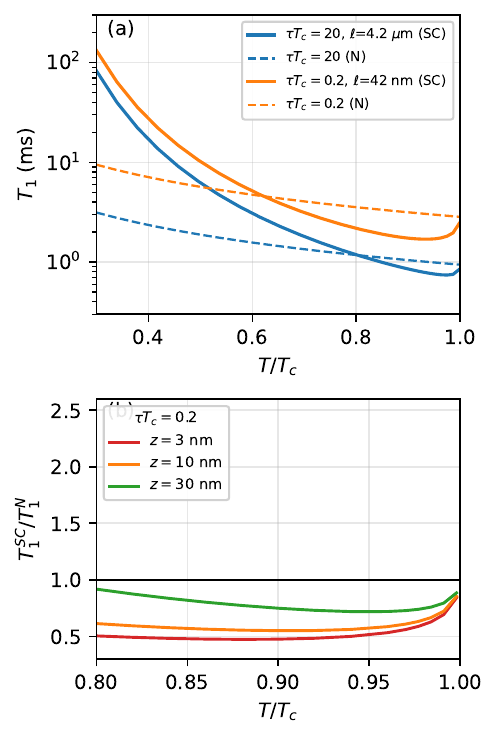}
\caption{(a) $T_1(T)$ for a spin qubit subject to magnetic noise from the surface of Nb in the superconducting (solid) and normal
(dashed) phases, without magnetic impurities, at $z=10$~nm for a clean ($\tau T_c=20$) and a dirty ($\tau T_c=0.2$) sample. (b) Ratio of the superconducting to normal-state $T_1$ just below $T_c$ for the dirty sample at three qubit heights. The coherence peak of $\sigma_1$ (Fig.~\ref{fig:sigma1}) produces a noise excess ($T_1^{SC}/T_1^{N}<1$) at all three heights, largest at the smallest height.}
\label{fig:t1B}
\end{figure}

The consequences of the above results for the $T$-dependence of $T_1$ in the absence of magnetic impurities are shown in Fig.~\ref{fig:t1B}(a) at $z=10$~nm for a clean ($\tau T_c=20$) and a dirty ($\tau T_c=0.2$) sample.  Two regimes are visible. Just below $T_c$ the superconductor is noisier than the normal metal: the coherence-factor enhancement of the dissipative conductivity (Fig.~\ref{fig:sigma1}) carries over to the noise, and $T_1^{SC}$ drops below $T_1^{N}$ down to $T\approx0.65\,T_c$ for $\tau T_c=0.2$ (down to $\approx0.8\,T_c$ for the very clean sample), with a maximal ratio $T_1^{SC}/T_1^{N}\simeq0.55$ at $z=10$~nm. At lower temperatures the superconducting $T_1$ increases exponentially above its normal-state value, proportional to $e^{\Delta/T}$: the gain is a factor $\approx1.8$ at $0.5\,T_c$, $\approx14$ at $0.3\,T_c$, and $\approx4\times10^{5}$ at $0.1\,T_c$. In contrast, the normal-state $T_1$ is nearly $T$-independent apart from the trivial $1/T_1 \propto \coth(\hbar\omega_0/2\kB T) \approx 2\kB T /\hbar\omega_0$ factor. The height dependence of the coherence peak is shown in Fig.~\ref{fig:t1B}(b): the noise excess just below $T_c$ is a factor $\approx2$ at $z=3$--$10$~nm ($T_1^{SC}/T_1^{N}\simeq0.48$ and $0.55$ at $0.9\,T_c$) and $\approx1.4$ at $z=30$~nm ($0.75$). The reduction with increasing $z$ is the result of two effects: (1) with increasing $z$ the dominant momenta $q\sim1/2z$ move from the nonlocal (Pippard) regime, where the coherence enhancement of $Q_2(q,\omega)$ is largest, toward the local regime; and (2) the superfluid contribution to $\re\,\eps_t$ screens the noise with increasing strength as $T$ drops; the screening parameter is $|{\re\,\eps_t}|/u^{2}$ at $u\sim1/2kz$, which grows as $z^{3}$ and scales as $\lambda^{-2}$ with the penetration depth. The latter dependence makes the visibility of the peak sensitive to the material: with the clean-limit London depth $c/\omega_p\approx10$~nm of a free-electron model of Nb the screening would be sixteen times stronger and the peak would be masked already at $z\approx10$~nm, whereas for the empirical $\lambda\approx40$~nm of Nb films it remains clearly visible at all typical NV working distances. We thus conclude that EWJN relaxometry of the Hebel-Slichter physics is feasible with shallow NV centers above Nb films, with the largest signal at the smallest heights. A more detailed discussion of the $z$-dependence of $T_1$ is given below; see also Fig.~\ref{fig:zdep}.

\begin{figure}[t]
\centering
\includegraphics[width=\columnwidth]{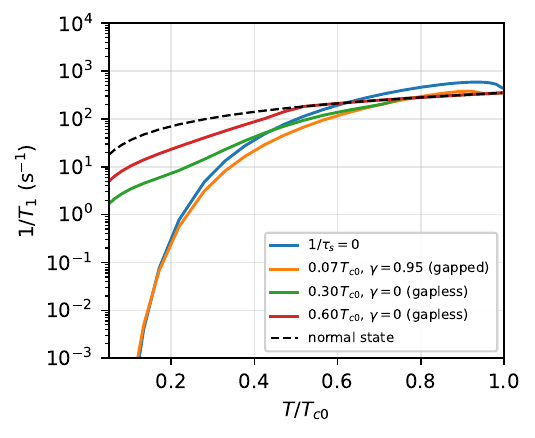}
\caption{Magnetic-impurity noise floor.  Relaxation rate $1/T_1(T)$ at $z=10$~nm,
$\tau T_c=0.2$, for increasing magnetic scattering; the normal-state rate is shown dashed.  Gapless
superconductivity converts the exponential drop of the rate into saturation at a level governed by $[\nu(0)/\nuF]^{2}$, Eq.~\eqref{eq:floorbound}. The vertical range is restricted to $T_1\lesssim10^{3}$~s; at longer times EWJN is irrelevant in practice because other relaxation mechanisms dominate.}
\label{fig:t1mag}
\end{figure}

We now discuss $T_1(T)$ in a superconductor with magnetic impurities. Our main result is reported in Fig.~\ref{fig:t1mag}, which shows the relaxation rate $1/T_1$. It is seen that, with increasing magnetic scattering, the low-temperature collapse of the rate is cut off and replaced by  saturation.  In the regime where the superconductor remains gapped ($1/\tau_s=0.07\,T_{c0}$, $\gamma=0.95$ in the figure), the curve tracks the clean one down to low $T$ (the residual DOS is exponentially small in this approximation; in reality optimal fluctuation tails and the Shiba band would set a floor at lower level). This is in sharp contrast to the gapless regime, where the emergence of a noise floor can be seen in Fig.~\ref{fig:t1mag}. For example, when
$1/\tau_s=0.30\,T_{c0}$, $\gamma=0$ (where $\nu(0)/\nuF\approx 0.45$), the qubit rate saturates near $1/T_1\approx1.7$~s$^{-1}$ ($T_1\approx 0.6$~s). 
In Appendix ~\ref{app:checks}, we show the general bound
    \begin{equation}
\frac{T_{1,N}(T)}{T_1(T)}\le\Big[\frac{\nu(0)}{\nuF}\Big]^2,\qquad T\ll T_c,
    \label{eq:floorbound}
    \end{equation}
where $T_{1,N}(T)$ is the relaxation time the same electrode would produce in its normal state at the same temperature (both rates carry the same Bose factor $\coth(\hbar\omega/2\kB T)$, so that the bound is a statement about the noise spectral densities). For the data shown in Fig.~\ref{fig:t1mag}, the bound is obeyed, with $T_1^{\rm floor}\,[\nu(0)/\nuF]^{2}/T_{1,N} \approx 2.1$ and $1.8$ for the two gapless cases shown in the figure; the factor above unity is the residual superfluid screening at $z=10$~nm. Finally, for strong pair breaking ($1/\tau_s=0.60\,T_{c0}$,
$T_c\approx 0.47\,T_{c0}$) the material is normal over much of the temperature range and the increase of $T_1$ as compared to the normal state never exceeds a factor of four. To be more precise, the noise (the quantity $\chi^{B}_\parallel$) becomes strictly temperature-independent at the floor.  The residual $T$-dependence of $1/T_1$ visible in Fig.~\ref{fig:t1mag} at $kT\gg\hbar\omega$ is the trivial $\coth$ factor and $1/T_1$ finally saturates at the quantum (spontaneous-emission) floor for $T\lesssim\hbar\omega/2\kB=70$~mK: gapless quasiparticles absorb the qubit energy even at $T=0$.

The saturation phenomenon we have described is the qubit-analog of the residual surface resistance of SRF
cavities \cite{proslier,kharitonov}: the same subgap states that limit cavity $Q$ at low temperature limit $T_1$ near superconducting electrodes.  Importantly, EWJN relaxometry constitutes a spatially resolved, contactless probe of $\nu(0)$ in the surface region of a superconductor.

\begin{figure}[t]
\centering
\includegraphics[width=\columnwidth]{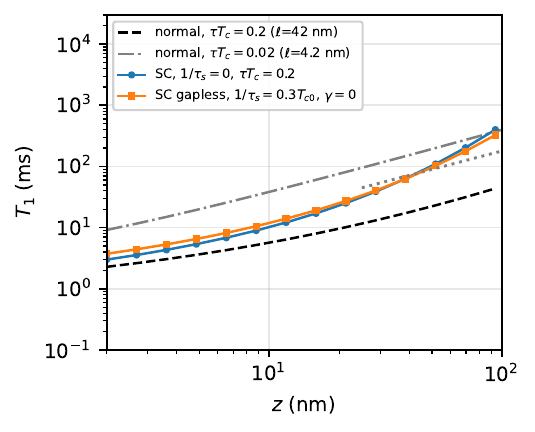}
\caption{Distance dependence of $T_1$ at $T=0.5\,T_{c0}$, for $2\,{\rm nm}\le z\le100$~nm. Normal state: $\tau T_c=0.2$ ($\ell=42$~nm, black dashed) and $\tau T_c=0.02$ ($\ell=4.2$~nm, gray dash-dotted); superconducting state with $\tau T_c=0.2$, without (circles) and with (squares) magnetic impurities. The dotted line (arrow) indicates the local near-field law $T_1\propto z$ [Eq.~\eqref{eq:Bpar-local}, valid for $\ell\ll z\ll\delta$]; the screened law $T_1\propto z^{4}$ [Eq.~\eqref{eq:z4law}] sets in beyond the plotted range.}
\label{fig:zdep}
\end{figure}

The height dependence of $T_1(z)$ is shown in Fig.~\ref{fig:zdep}  at $T=0.5\,T_{c0}$. Over the range $2$--$100$~nm none of the curves is a pure power law, because this range straddles the material length scales. For the normal metal with $\tau T_c=0.2$ the mean free path $\ell=42$~nm lies inside the plotted range, so the noise crosses over from the anomalous (nonlocal) regime at $z\lesssim\ell$---where the dissipation is Landau damping with $\im\,\eps_t\propto1/q^{3}$, which weakens the $z$-dependence---to the local regime: the local exponent ${\rm d}\ln T_1/{\rm d}\ln z$ rises from $\approx0.5$ at $z=2$~nm to $\approx1$ at $z\approx50$~nm. The local near-field law $T_1\propto z$ of Eq.~\eqref{eq:Bpar-local} requires $\ell\ll z\ll\delta$ (here $\delta\approx1\,\mu$m); it is displayed over most of the range by the dirtier sample with $\tau T_c=0.02$ ($\ell=4.2$~nm, $\delta\approx3\,\mu$m), for which the exponent is $\approx1$ for $z\gtrsim10$~nm. The superconducting curve is steeper. At the smallest heights the dominant partial waves are not screened ($u^2\gg|\re\,\eps_t|$) and the superconductor is only modestly quieter than the normal metal at this temperature ($T_1^{SC}/T_1^{N}\approx1.3$ at $z=2$~nm), because the nonlocal kernel $\propto1/q$ of the Pippard regime gives a nearly $z$-independent noise and the coherence-factor enhancement partly compensates the quasiparticle depletion. Once $|\re\,\eps_t|\gtrsim u^{2}$ at $u\sim1/2kz$, i.e., for $z\gtrsim\lambda/2\approx20$~nm, London screening of the fluctuating currents takes over and the exponent grows continuously ($1.4$ at $20$~nm, $2.4$ at $100$~nm), approaching the steep law $T_1\propto z^{4}$ derived in Appendix~\ref{app:half_space_greens_func} [Eq.~\eqref{eq:z4law}] only for $z\gtrsim0.5\,\mu$m; the superconducting advantage grows accordingly, from $1.3$ at $2$~nm to $\approx6$ at $70$~nm. The gapless curve differs from the clean superconducting one by less than $25\%$ at this temperature, because at $T=0.5\,T_{c0}$ the thermally excited quasiparticles still dominate the dissipation; the two curves separate only at low temperature, where the residual density of states sets the floor of Fig.~\ref{fig:t1mag}.

\subsection{Electric noise and charge-qubits}
\label{sec:electric}

We now consider the case of electric noise and the corresponding relaxation rates of nearby charge-qubits. For a charge qubit with dipole matrix element $d$ along $\hat z$, we have 
    \begin{equation}
    \frac{1}{T_1}= \frac{d^2}{\hbar ^2}\chi_\perp^{E}(\mathbf r_0, \omega_0)\coth\left(\frac{\hbar \omega_0}{2k_BT}\right),
    \end{equation}
where $\chi^E$ is the electric-field response function (cf. Eq.~\eqref{eq:T1_chi}) and we similarly define the dimensionless noise strength $b_{\parallel/\perp}^E(z,\omega) \equiv (4\hbar \omega^3 / 3 c^3)^{-1} \chi_{\parallel/\perp}^E(z,\omega)$. Formulas for $b_{\parallel/\perp}^E$ are obtained from the corresponding magnetic noise formulas \eqref{eq:bpar} and \eqref{eq:bperp} by swapping $r_s \leftrightarrow r_p$.

Electric noise is governed by $\im~r_p$ (see Eq.~\eqref{eq:bperp} with $r_s \leftrightarrow r_p$) and, in the near field, by the electrostatic term of \eqref{eq:zetap}, where we can safely replace $r_p \to (\epsilon_l - 1)/(\epsilon_l + 1)$ evaluated at $q \approx 1/2z$. Combining this with the $w$-integral of Eq.~\eqref{eq:bperp} gives the familiar $1/z^{3}$ law of near-field electric noise, $b^E_\perp\simeq\tfrac38(kz)^{-3}\,\im[(\bar\eps_l-1)/(\bar\eps_l+1)]$, with $\bar\eps_l=\eps_l(q\simeq1/2z,\omega)$. Two consequences follow immediately. First, the electric noise is controlled by the longitudinal (density) response, so that the coherence factors of the charge-density vertex enter. Second, since $\re\,\eps_l\simeq1+q_{\rm TF}^2/q^2$ is dominated by Thomas-Fermi screening and is essentially unaffected by superconductivity, $\im[(\eps_l-1)/(\eps_l+1)]\simeq2\,\im\,\eps_l/|\eps_l+1|^{2}$ is proportional to the dissipative part of the longitudinal response alone.

To evaluate $\im\,\eps_l$ below $T_c$ we use the following approximation, which we state together with its justification. The charge density is even under time reversal, so that in BCS theory its coherence factors are of Tinkham's case I, $(EE'-\Delta^{2})/EE'$ \cite{tinkham}---the combination that suppresses quasiparticle scattering at $\hbar\omega\ll\Delta$, in contrast to the case-II factors $(EE'+\Delta^2)/EE'$ of the current vertex in Eq.~\eqref{eq:cohfactors}. For $\hbar\omega\ll\Delta$ the dissipative part of the density response is a quasiparticle-scattering process weighted by $-\partial n_F/\partial\ee$, and its ratio to the normal-state value is (Appendix~\ref{app:caseI})
\begin{equation}
R_{I}(T)
=\int \dd\ee\Big(-\frac{\partial n_F}{\partial\ee}\Big)
\Big[\big(\re\, g\big)^{2}-\big(\re\, f\big)^{2}\Big],
\label{eq:caseI}
\end{equation}
where $R_I\equiv\im\,\eps_l^{SC}/\im\,\eps_l^{N}$, and $g$ and $f$ are the disorder-averaged functions of Sec.~\ref{sec:results}. In the clean limit $(\re g)^2-(\re f)^2=1$ above the gap and $0$ below it, so that $R_I=2n_F(\Delta)$, the textbook result for the ultrasonic attenuation ratio \cite{tinkham}; in the gapless regime $R_I(0)=[\nu(0)/\nuF]^2$, i.e., the charge noise acquires the same type of residual floor as the magnetic noise. We therefore model the longitudinal function below $T_c$ as $\eps_l^{SC}(q,\omega)=\re\,\eps_l^{N}(q,\omega)+\ii R_I(T)\,\im\,\eps_l^{N}(q,\omega)$, with $\eps_l^{N}$ the Mermin function \eqref{eq:mermin2}. This construction is exact in the local dirty limit and captures the coherence-factor physics and the correct normal-state limit in general; it neglects the modification of the $q$-dependence of the dissipative density response by pairing in the nonlocal regime, whose consistent treatment requires the longitudinal BCS kernel with conserving (charge-vertex) corrections and is beyond the scope of this work. The transverse term of $\zeta_p$ in Eq.~\eqref{eq:zetap} is evaluated with the full kernel \eqref{eq:kernelfinal}.

\begin{figure}[t]
\centering
\includegraphics[width=\columnwidth]{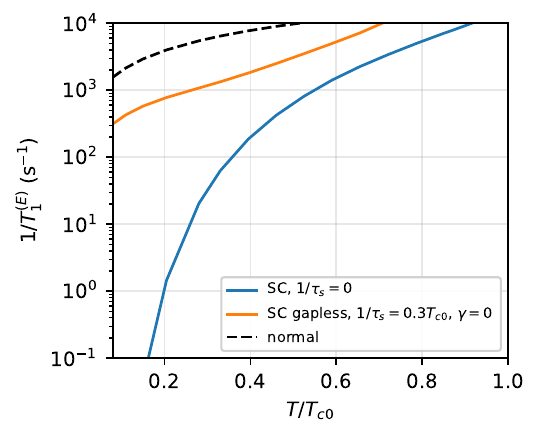}
\caption{Charge-qubit relaxation rate $1/T_1^{(E)}$ from electric EWJN above Nb for a dipole moment $d=ea_0$ oriented along the surface normal, at $z=10$~nm and $\tau T_c=0.2$, in the normal state (dashed), the superconducting state without magnetic impurities, and the gapless state. The vertical range is restricted to $T_1^{(E)}\lesssim10$~s.}
\label{fig:t1E}
\end{figure}

Figure~\ref{fig:t1E} shows the resulting charge-qubit relaxation rate for a dipole $d=ea_0$ ($a_0$ the Bohr radius) at $z=10$~nm. As anticipated: (i) the noise drops steeply below $T_c$---more steeply than the magnetic noise, since the dissipative channel is suppressed while the screening denominator $|\eps_l+1|^2$ is not; (ii) there is no Hebel--Slichter feature, in accordance with the case-I coherence factors; and (iii) gapless superconductivity again imposes a temperature-independent floor. We caution that for charge qubits at these distances the intrinsic EWJN electric noise competes with extrinsic $1/f$ charge noise, and that the absolute floor in the gapless case inherits the approximation made for $\eps_l^{SC}$; the qualitative conclusions are robust, whereas the absolute numbers are less certain than in the magnetic case.

\section{Experimental implications and outlook}
\label{sec:conclusion}

We have developed a theory of EWJN above bulk superconductors, accounting for the possibility of both non-magnetic and magnetic impurities. The theory is built around  the FDT/electromagnetic Green's function formalism (Sec.~\ref{sec:formalism} and Appendix~\ref{app:half_space_greens_func}), the specular-reflection nonlocal surface impedances (Appendix~\ref{app:nonlocal_impedances}), and the microscopic Shiba-theory current kernel, which contains Drude, Boltzmann, London, Pippard,
Mattis-Bardeen, Abrikosov-Gor'kov, and gapless physics in the appropriate limits (Sec.~\ref{sec:results} and Appendix~\ref{app:checks}). The primary results are the noise cliff below $T_c$; the case-II coherence
peak in magnetic noise (a factor-of-two noise excess just below $T_c$ at typical NV heights, reduced with increasing height by superfluid screening) and its absence
in electric noise (case I); and the magnetic-impurity noise floor,
Eqs.~\eqref{eq:gaplessfloor}--\eqref{eq:floorbound}, the qubit analog of
residual SRF cavity resistance.

There are a number of connections between our work and current, as well as potentially future, experimental investigations on superconducting materials. NV relaxometry over superconducting films has recently been carried out in Refs.~\cite{liu2025,li2026}. Replacing the silver film of the Kolkowitz \emph{et al.}\ experiment \cite{kolkowitz2015} by Nb (or Pb) and sweeping $T$ through $T_c$ at working distances $z=5$--$50$~nm would map the entire physics computed here: the cliff, the coherence peak (a reduction of $T_1$ by a factor $\approx2$ just below $T_c$ at $z=5$--$10$~nm, decreasing to $\approx30\%$ at $z=30$~nm), and the low-temperature floor.  Near $T_c$, $\dd T_1/\dd T$ is enormous, making the NV a sensitive local thermometer; deep in the superconducting state, $T_1$ directly reads the subgap density of states.

We have also demonstrated the utility of noise sensing for magnetic-impurity diagnostics. For example, controlled surface magnetic
doping (or the native magnetic moments of the Nb$_2$O$_5$ suboxide layer \cite{proslier}) should produce the floor of Fig.~\ref{fig:t1mag} with $T_1^{\rm floor}$ tracking $[\nu(0)/\nuF]^{-2}$.  Since the same states produce the residual SRF resistance, a quantitative cross-check between
cavity $Q_0$ measurements and near-surface qubit relaxometry on identically prepared material becomes possible.

Our results also provide design guidance for hybrid SC devices. For spin qubits near superconducting gates, the benefit of superconducting electrodes saturates at $T_1^{\rm floor}=\alpha[\nuF/\nu(0)]^{2}\,T_1^{\rm normal}$, where $\alpha\ge1$ is a screening factor ($\alpha\approx2$ at $z=10$~nm for Nb).  Keeping the EWJN advantage beyond $10^{4}$ requires $\nu(0)/\nuF\lesssim10^{-2}$, i.e.\ magnetically clean surfaces; conversely there is little to gain from cooling below the temperature where the floor is reached.

There are a number of natural extensions of the work presented here. The formalism for noise in terms of reflection coefficients $r_{s,p}$ is readily generalized to describe thin films and realistic gate geometries. In the diffusive limit, the connection between our results and the Usadel approach of Ref.~\cite{fominov} is well worth investigating. An exact longitudinal (density) kernel with conserving vertex corrections, improving the results of  Sec.~\ref{sec:electric}, needs to be developed. We have not considered the effects of EWJN on dephasing ($T_2$, $T_\varphi$), which can arise from the low-frequency tail of the noise spectral densities and could be relevant to experimental protocols. Our treatment of disorder can also be extended beyond the non-crossing approximation to include the DOS tails \cite{lamacraft}, which will soften the hard-gap results at very low $T$. In this paper we have not investigated the two-point correlations of the noise, which can give additional information about the system~\cite{premakumar2018,rovny2025}. Finally, it is of significant interest to investigate situations with nonequilibrium quasiparticle distributions, which are highly relevant for current experiments on SC qubits; a residual fractional quasiparticle density $x_{\rm qp}$ enters Eq.~\eqref{eq:kernelfinal} through the distribution function and could be expected to produce a floor analogous to the magnetic-impurity one.

\begin{acknowledgments}
This work was supported by the U.S. Department of Energy (DOE), Office of Science, Basic Energy Sciences (BES) under Award No. DE-SC0020313.
A. L. acknowledges H. I. Romnes Faculty Fellowship provided by the University of Wisconsin-Madison Office of the Vice Chancellor for Research and Graduate Education with funding from the Wisconsin Alumni Research Foundation.
G.R. acknowledges financial support from the Sweden-America Foundation through the Ingegerd \& Viking Olov Bj\"orks stipendiefond.
The authors acknowledge the use of the large language model Claude (Anthropic) in the preparation of the manuscript, including, in particular, numerical calculations, graphics, and symbolic verification of analytical results. The authors conceived the project and carried out the research.
\end{acknowledgments}

\appendix

\section{Half-space Green's functions}
\label{app:half_space_greens_func}

In this Appendix we derive Eqs.~\eqref{eq:bpar} and \eqref{eq:bperp} and the local asymptotic forms used in the main text. We use Gaussian units, time dependence $e^{-\ii\omega t}$, and $k=\omega/c$. The material occupies $z<0$ and the qubit sits on the $\hat z$ axis at height $z>0$.

\emph{Free-space dyadic Green's function.}  The electric field of a point electric dipole $\bm p$ at the origin obeys $\nabla\times\nabla\times\bm E-k^2\bm E=4\pi k^2\bm p\,\delta(\bm r)$, so that $\bm E=\mathsf G^0\bm p$ with $\mathsf G^0(\bm r)=(\nabla\nabla+k^2\mathsf 1)\,e^{\ii kr}/r$. The Weyl identity \cite{weyl1919}
\begin{equation}
\frac{e^{\ii kr}}{r}=\frac{\ii}{2\pi}\int\dd^2Q\,\frac{e^{\ii\bm Q\cdot\bm\rho}e^{\ii k_z|z|}}{k_z},
\label{eq:weyl}
\end{equation}
with $k_z=\sqrt{k^2-Q^2}$, $\im k_z\ge0$, expands the field in plane waves labeled by the lateral wave vector $\bm Q$; in the dimensionless variables $u=Q/k$ and $v=\sqrt{1-u^2}$ (${\rm Im}\,v\ge0$) one has $k_z=kv$, and $u<1$ ($u>1$) labels propagating (evanescent) waves. Taking the imaginary part of $\mathsf G^0$ at coincident points gives $\im\mathsf G^0_{ii}=\tfrac23k^3$, to which only propagating waves contribute.

\emph{Reflected part.}  Each plane-wave component of the dipole field is reflected by the surface with the coefficient $r_s(u)$ or $r_p(u)$ appropriate to its polarization, $\hat{\bm e}_s=\hat{\bm Q}\times\hat{\bm z}$ and $\hat{\bm e}^{\pm}_p=(Q\hat{\bm z}\mp k_z\hat{\bm Q})/k$ for waves traveling toward $\pm z$, and acquires the phase $e^{2\ii k_zz}$ for the round trip from the source to the surface and back. Reassembling the dyadic at the source point and performing the azimuthal average ($\cos^2\phi,\sin^2\phi\to\tfrac12$) yields \cite{wyliesipe,novotny}
\begin{align}
\mathsf G^{\rm scat}_{xx}=\mathsf G^{\rm scat}_{yy}&=\frac{\ii k^3}{2}\int_0^\infty\frac{u\,\dd u}{v}\big[r_s(u)-v^2r_p(u)\big]e^{2\ii kvz},
\label{eq:Gxx}\\
\mathsf G^{\rm scat}_{zz}&=\ii k^3\int_0^\infty\frac{u^3\dd u}{v}\,r_p(u)\,e^{2\ii kvz}.
\label{eq:Gzz}
\end{align}
The $s$ contribution to the $xx$ component carries $\hat e_{s,x}^2=\sin^2\phi$, the $p$ contribution $\hat e^{+}_{p,x}\hat e^{-}_{p,x}=-(k_z^2/k^2)\cos^2\phi$, and only $p$ waves, with weight $(Q/k)^2$, contribute to $zz$. Inserted into Eq.~\eqref{eq:T1master} at $T=0$, Eqs.~\eqref{eq:Gxx}--\eqref{eq:Gzz} reproduce the classical Chance--Prock--Silbey formulas \cite{cps} for the modified spontaneous-emission rate near an interface, $\Gamma_\perp/\Gamma_0=1+\tfrac32\re\int_0^\infty(u^3/v)\,r_p\,e^{2\ii kvz}\dd u$ and $\Gamma_\parallel/\Gamma_0=1+\tfrac34\re\int_0^\infty(u/v)[r_s-v^2r_p]e^{2\ii kvz}\dd u$, a check of all prefactors.

\emph{Magnetic Green's function.}  The magnetic field of a magnetic dipole obeys the same wave equation, and at the surface the roles of the two polarizations are interchanged ($s$-polarized electric waves are $p$-polarized magnetic waves and vice versa). Hence $\mathsf G^{B,\rm scat}$ is obtained from Eqs.~\eqref{eq:Gxx}--\eqref{eq:Gzz} by the substitution $r_s\leftrightarrow r_p$.

\emph{Evanescent parametrization.}  For $u>1$ we write $u=\sqrt{1+w^2}$, $v=\ii w$, so that $u\,\dd u=w\,\dd w$ and $e^{2\ii kvz}=e^{-2kzw}$. Then $\re[(u/v)(r_p-v^2r_s)e^{2\ii kvz}]\dd u=e^{-2kzw}\,\im[r_p+w^2r_s]\dd w$ and $\re[(u^3/v)r_se^{2\ii kvz}]\dd u=(1+w^2)e^{-2kzw}\im r_s\,\dd w$, which, together with $\chi_{ij}=2\hbar\im G_{ij}$ and the definition $b\equiv(4\hbar\omega^3/3c^3)^{-1}\chi$, give Eqs.~\eqref{eq:bpar}--\eqref{eq:bperp}. In the near field, $z\ll c/\omega$, the evanescent integrals are of order $(kz)^{-1}$ to $(kz)^{-3}$ times $\im\eps$ and therefore exceed the vacuum and propagating ($u<1$) contributions, which are of order unity, by many orders of magnitude; the latter are dropped. Passivity requires $\im r_{s,p}>0$ for evanescent waves, so that the $b$'s are positive.

\emph{Local asymptotics.}  For a local good conductor, $|\eps|\gg1$, three regimes of Eq.~\eqref{eq:bpar} are useful. (a) \emph{Electric noise}, $z\ll\delta$: for $u\gg\sqrt{|\eps|}$, $r_p\to(\eps-1)/(\eps+1)$ and $b^E_\perp\simeq\tfrac32\im[(\eps-1)/(\eps+1)]\int_0^\infty w^2e^{-2kzw}\dd w=\tfrac38(kz)^{-3}\im[(\eps-1)/(\eps+1)]$. (b) \emph{Magnetic noise}, $\ell\ll z\ll\delta$: with $s=\sqrt{\eps-u^2}\simeq\ii u(1-\eps/2u^2)$ for $u^2\gg|\eps|$ one finds $\im r_s\simeq\im\eps/4u^2$, and the $w^2r_s$ term of Eq.~\eqref{eq:bpar} gives Eq.~\eqref{eq:Bpar-local}. (c) \emph{Magnetic noise in the screened regime}, $u^2\ll|\eps|$ (which is the dominant range when $z\gg\delta$, or, in a superconductor, when $z$ exceeds the London scale discussed in Sec.~\ref{subsec:magnetic}): writing $s\simeq\sqrt\eps$ and expanding $r_s=(v-s)/(v+s)$ to first order in $\im\eps$, ${\rm d}r_s/{\rm d}\eps=w/[S(w+S)^2]$ with $S=\sqrt{|\eps|}$, so that $\im r_s\simeq w\,\im\eps/|\eps|^{3/2}$ for $w\ll S$. Then
\begin{equation}
b^B_\parallel\simeq\frac34\,\frac{\im\eps}{|\eps|^{3/2}}\int_0^\infty w^3e^{-2kzw}\dd w
=\frac94\,\frac{\im\eps}{|\eps|^{3/2}}\,\frac{1}{(kz)^4},
\label{eq:z4law}
\end{equation}
i.e., $T_1\propto z^4$: the fluctuating currents that dominate at large $z$ are screened by the medium. For a superconductor $|\eps_t|\simeq c^2/\lambda^2\omega^2$ at the relevant $q$, and the crossover between regimes (b) and (c) occurs when $u\sim1/2kz$ equals $\sqrt{|\eps_t|}=c/\lambda\omega$, i.e., at $z\sim\lambda/2$, with $\lambda$ the effective penetration depth at the relevant wave vector ($\approx20$~nm for Nb).

\section{Nonlocal surface impedances}
\label{app:nonlocal_impedances}

Here we derive Eqs.~\eqref{eq:zetas} and \eqref{eq:zetap} in the specular-reflection (semiclassical infinite barrier) model \cite{reuter,kliewer1968,fordweber,Gerhardts1983}. The medium fills $z<0$; fields vary as $e^{\ii Qx-\ii\omega t}$ along the surface.

\emph{Specular reflection as even extension.}  If electrons are reflected specularly, the distribution function and hence the current inside the medium are the same as in an infinite medium in which the fields have been extended as even functions of $z$ about the surface. The half-space problem with a nonlocal bulk kernel therefore maps onto a translationally invariant one, with the surface entering only through the discontinuity of the derivative of the extended field at $z=0$.

\emph{$s$ polarization.}  Let $E=E_y(z)$. Inside the medium Maxwell's equations give $E_y''+(\omega^2/c^2)\hat\eps_tE_y-Q^2E_y=0$, where $\hat\eps_t$ acts by convolution in $z$. The even extension $\bar E_y(z)=E_y(-|z|)$ satisfies the bulk equation with a surface source, $\bar E_y''=E_y''+2E_y'(0^-)\delta(z)$, so that in Fourier space ($\bar E_y(z)=\int\frac{\dd q}{2\pi}e^{\ii qz}\tilde E(q)$)
\begin{equation}
\Big[\frac{\omega^2}{c^2}\eps_t\big(\sqrt{q^2+Q^2},\omega\big)-q^2-Q^2\Big]\tilde E(q)=2E_y'(0^-),
\end{equation}
and hence
\begin{equation}
\frac{E_y(0)}{E_y'(0^-)}=\int_{-\infty}^{\infty}\frac{\dd q}{\pi}\,\frac{1}{(\omega^2/c^2)\eps_t-q^2-Q^2}.
\end{equation}
The surface impedance of this polarization is $Z_s=E_\parallel(0)/H_\parallel(0)$ with $H_x=-(\ii c/\omega)E_y'$. In the dimensionless variables $u=cQ/\omega$, $y=cq/\omega$, $\kappa^2=u^2+y^2$ this gives $Z_s=\zeta_s/\pi$ with $\zeta_s$ of Eq.~\eqref{eq:zetas}. On the vacuum side the impedance of a partial wave with lateral momentum $u$ is $1/v$, and matching the tangential fields yields the reflection coefficient Eq.~\eqref{eq:rs}.

\emph{$p$ polarization.}  For TM waves both $E_x$ and $E_z$ are present and $\nabla\cdot\bm E\ne0$ inside the medium, so that the longitudinal response participates. Applying the same even-extension procedure to the pair $(E_x,E_z)$, using $\nabla\cdot\bm D=0$ to eliminate $E_z$, and separating the Fourier components of the field into parts transverse and longitudinal to the three-dimensional wave vector $(Q,0,q)$, one finds \cite{fordweber} the impedance $Z_p=\zeta_p/\pi$ with $\zeta_p$ of Eq.~\eqref{eq:zetap}; the vacuum-side impedance of a $p$ wave is $v$, and matching gives Eq.~\eqref{eq:rp}. The two terms of Eq.~\eqref{eq:zetap} have a transparent meaning: the $y^2$ term is the penetrating (transverse) electromagnetic wave, while the $u^2/\eps_l$ term is the electrostatic (longitudinal) response. At large $u$ the latter dominates and $r_p$ reduces to the image-charge form $(\eps_l-1)/(\eps_l+1)$ with $\eps_l$ evaluated at $q\simeq(\omega/c)u$.

\emph{Local limit.}  For $q$-independent $\eps$ the integrals are elementary. With $s=\sqrt{\eps-u^2}$ ($\im s>0$), $\int_0^\infty\dd y/(\eps-u^2-y^2)=\ii\pi/2s$ gives $\zeta_s=\pi/s$, and the partial-fraction identity $y^2/[\kappa^2(\eps-\kappa^2)]+u^2/(\kappa^2\eps)=(\eps-u^2)/[\eps(\eps-\kappa^2)]$ gives $\zeta_p=\pi s/\eps$; Eqs.~\eqref{eq:rs} and \eqref{eq:rp} then reduce to the Fresnel coefficients \eqref{eq:fresnel}, which we have also verified numerically to better than $10^{-6}$.

\emph{Validity.}  The response functions used in this work are computed in the quasiclassical (Fermi-surface) approximation and hold for $q\ll k_F$. Since the noise integrals are cut off at $q\sim1/2z$ by the factor $e^{-2kzw}$, all results are reliable for $z\gtrsim$ a few times $k_F^{-1}\approx0.1$~nm, comfortably covering the experimentally relevant range $z\gtrsim2$~nm. At sub-nanometer distances additional physics (band structure, electron spill-out, the Landau-damping cutoff at $q>2k_F$) would enter and is not considered here.

\section{Limiting behaviors of the transverse conductivity and relation to earlier theories}
\label{app:checks}

In this Appendix we show that the kernel \eqref{eq:kernelfinal} reproduces the established forms of the transverse response in all relevant limits, and we derive the residual dissipation in the gapless regime. Each statement has also been verified numerically (Appendix~\ref{app:numerics}).

\emph{1. Normal state.}  For $\Delta\to0$ one has $g\to1$, $f\to0$, $h\to\ee$, $C_{RA}\to1$, $C_{RR}\to0$, and $\int\dd\ee\,(\thf_+-\thf_-)=2\omega$; Eq.~\eqref{eq:kernelfinal} collapses to the Boltzmann kernel \eqref{eq:QN}, and at $k\to0$ to the Drude form $\hat Q=\omega/(\omega+\ii/\tau)$.

\emph{2. London limit.}  For a clean system at $T\to0$, $k\to0$, and $\omega\ll\Delta$ the kernel is real and $\hat Q\to1$, i.e., $\eps_t\simeq-\omega_p^2/\omega^2$ and $\lambda(T=0)=\lambda_{L0}$. At finite $T$, $\hat Q(0,0)=n_s(T)/n$ defines the superfluid fraction, and the two-fluid forms $\sigma_2\simeq n_se^2/m\omega$, $\sigma_1\simeq n_ne^2\tau/m$ emerge from Eq.~\eqref{eq:kernelfinal} in the appropriate expansion.

\emph{3. Pippard (extreme anomalous) limit.}  For a clean system at $T\to0$ and $\vF k\gg\Delta$, all $\mathcal A\to\pi/4\vF k$ and the energy integral gives
\begin{equation}
\hat Q(k,0)\to\frac{3\pi^2}{4}\,\frac{\Delta}{\vF k},
\label{eq:pippard}
\end{equation}
the BCS nonlocal kernel \cite{tinkham}. Our numerics approaches Eq.~\eqref{eq:pippard} from below with the expected slow $O[\ln(\vF k/\Delta)\,\Delta/\vF k]$ corrections (ratio $0.96$ at $\vF k/\Delta=1900$).

\emph{4. Dirty local limit: Mattis--Bardeen and the coherence peak.}  For $1/\tau\gg\Delta$ and $k\to0$, expanding $\mathcal A$ at large $l_0$ turns Eq.~\eqref{eq:kernelfinal} into the Mattis--Bardeen conductivity \cite{mattisbardeen}. In particular, for $\hbar\omega<2\Delta$ the dissipative part is
\begin{align}
\frac{\sigma_1}{\sigma_n}
&=\frac{2}{\omega}\int_{\Delta}^{\infty}\dd E\,
\big[n_F(E)-n_F(E+\omega)\big]\nonumber\\
&\qquad\times\frac{E(E+\omega)+\Delta^{2}}
{\sqrt{E^{2}-\Delta^{2}}\sqrt{(E+\omega)^{2}-\Delta^{2}}},
\label{eq:MB}
\end{align}
where $\sigma_1=Q_2/\omega$ and $Q=Q_1-\ii Q_2$. Our kernel reproduces Eq.~\eqref{eq:MB} to better than $10^{-3}$ over $0.3\le T/T_c\le0.95$. The coherence factor $(EE'+\Delta^2)$ in Eq.~\eqref{eq:MB} is of Tinkham's case II \cite{tinkham}: low-frequency electromagnetic absorption is a quasiparticle-\emph{scattering} process, like nuclear-spin relaxation, not a pair-breaking one. Consequently $\sigma_1/\sigma_n$ has a Hebel--Slichter-type peak just below $T_c$: for $\omega\to0$ the integral \eqref{eq:MB} diverges logarithmically at $E=\Delta$ and is cut off by $\omega$ itself,
\begin{equation}
\frac{\sigma_1}{\sigma_n}\simeq\frac{\Delta}{2T}\,\frac{1}{\cosh^2(\Delta/2T)}\,\ln\frac{c_1\Delta}{\omega}+\ldots,
\label{eq:HSasympt}
\end{equation}
with $c_1$ a constant of order unity, producing a maximum $\simeq2.5$ at $T\simeq0.85\,T_c$ for $\omega/2\pi=2.88$~GHz in Nb (Fig.~\ref{fig:sigma1}). By contrast, the longitudinal (density) channel carries case-I factors $(EE'-\Delta^2)$ and shows no peak; this asymmetry is what distinguishes magnetic from electric EWJN in Sec.~\ref{sec:electric} and Appendix~\ref{app:caseI}.

\emph{5. Gapless dissipation at $T\to0$.}  For $\omega\ll\Delta,T$ the dissipative part of Eq.~\eqref{eq:kernelfinal} can be written as $Q_2(\omega,k)=Q_0\,\omega\int\dd\ee\,(-\partial_\ee n_F)\,\bar Q_2(\ee,k)$ with
\begin{align}
\bar Q_2(\ee,k)&=\frac32\,\re\Big\{\big[1+|g|^2+|f|^2\big]\mathcal A(l_0^{RA})\nonumber\\
&\qquad+f^2\mathcal A(l_0^{RR})+f^{*2}\mathcal A^{*}(l_0^{RR})\Big\}_{\omega\to0},
\label{eq:Q2bar}
\end{align}
which follows from expanding $\thf_\pm$ to first order in $\omega$ and using $g^2-f^2=1$ (note that $C_{RR}\to-f^2$ at $\omega=0$, and that $Q_2=-\im Q=\tfrac32\re\int(\cdots)$ because of the prefactor $-3\ii/2$ in Eq.~\eqref{eq:kernelfinal}). Equation~\eqref{eq:Q2bar} has the property $\bar Q_2(\ee,k)=0\Leftrightarrow\nu(\ee)=0$: wherever $g$, $f$, and $h$ are simultaneously imaginary (the gapped region; see the remark below Eq.~\eqref{eq:hcorrect}) the three $\mathcal A$ functions coincide, and writing $g=-\ii g_2$, $f=-\ii f_2$ the coherence factors combine to $1+g_2^2-f_2^2=0$, so that dissipation vanishes identically. Conversely, at $\ee=0$ in the gapless regime $g(0)=g_1(0)$ is real while $f(0)$ and $h(0)$ are imaginary; the three $\mathcal A$'s are again equal and real, $|f(0)|^2=1-g_1(0)^2$, and the coherence combination gives
\begin{equation}
\bar Q_2(0,k)=3\,g_1(0)^{2}\,\mathcal A(l_0),\qquad l_0=\frac{2h(0)+\ii/\tau}{\vF k},
\label{eq:gaplessfloor}
\end{equation}
with $g_1(0)=\nu(0)/\nuF$.
The zero-temperature dissipation is thus proportional to the \emph{square} of the residual density of states. In the extreme nonlocal regime $\vF k\gg|h(0)|,1/\tau$, where $\mathcal A\to\pi/4\vF k$ takes the same value as in the normal state, Eq.~\eqref{eq:gaplessfloor} implies the simple law quoted in the abstract: the residual noise kernel is $[\nu(0)/\nuF]^2$ times the normal-state one. At smaller $k$ (larger $z$) the superfluid contribution to $\re\,\eps_t$ additionally screens the noise, so that in general
\begin{equation}
\frac{T_{1,N}(T)}{T_1(T)}\le\Big[\frac{\nu(0)}{\nuF}\Big]^2,\qquad T\ll T_c,
\end{equation}
with equality approached at small $z$, which is the result quoted in Eq.~\eqref{eq:floorbound} of the main text; $T_{1,N}(T)$ is the normal-state value at the same temperature, so that the common Bose factor cancels in the ratio. Numerically (Nb, $z=10$~nm, $\tau T_c=0.2$) the floor sits a factor $\approx2$ below the bound (Sec.~\ref{subsec:magnetic}).

\emph{Relation to earlier theories.}  Setting $1/\tau_s=0$ in Eq.~\eqref{eq:kernelfinal} yields the disordered BCS transverse response at arbitrary $(q,\omega,T)$---the content of Nam's theory \cite{nam1967a,*nam1967b} and, at $q=0$, of Mattis--Bardeen \cite{mattisbardeen} and Abrikosov--Gor'kov--Khalatnikov \cite{agk}; the present formulation is more compact and better suited to numerics (a single energy integral of algebraic functions). Setting $\Delta=0$ gives the nonlocal normal-state response of Sec.~\ref{subsec:normal_state_response}. The kernel is quasiclassical and particle--hole symmetric; it does not contain the $q\gtrsim k_F$ cutoff of the exact Lindhard response, which is irrelevant at $z\gg k_F^{-1}$ (Appendix~\ref{app:nonlocal_impedances}).

\section{Coherence factors of the longitudinal channel}
\label{app:caseI}

We derive Eq.~\eqref{eq:caseI}. For a perturbation $\hat O$ coupling to quasiparticles, the low-frequency absorption due to scattering of thermally excited quasiparticles is proportional to $\int\dd\ee\,(-\partial_\ee n_F)\,[N(\ee)^2\pm M(\ee)^2]$, where $N(\ee)=\nuF\re g(\ee)$ is the density of states, $M(\ee)=\nuF\re f(\ee)$ is the anomalous (pairing) density, and the sign is $+$ for operators odd under time reversal (case II: current, spin) and $-$ for operators even under time reversal (case I: density) \cite{tinkham}. This is the disorder-averaged form of the familiar BCS coherence factors $(EE'\pm\Delta^2)/EE'$, to which it reduces in the clean limit where $\re g=E/\sqrt{E^2-\Delta^2}$ and $\re f=\Delta/\sqrt{E^2-\Delta^2}$ above the gap. For the density (longitudinal) channel the ratio to the normal state, in which $N=\nuF$ and $M=0$, is therefore
\begin{equation}
R_I(T)=\int\dd\ee\Big(-\frac{\partial n_F}{\partial\ee}\Big)\Big[(\re g)^2-(\re f)^2\Big],
\end{equation}
which is Eq.~\eqref{eq:caseI}. In the clean limit $(\re g)^2-(\re f)^2=(E^2-\Delta^2)/(E^2-\Delta^2)=1$ for $|\ee|>\Delta$ and $0$ otherwise, so that $R_I=\int_{|\ee|>\Delta}(-\partial_\ee n_F)\dd\ee=2n_F(\Delta)$, the classic result for the ultrasonic attenuation ratio \cite{tinkham}, which decreases monotonically below $T_c$. In the gapless regime, at $T\to0$ only $\ee=0$ contributes, where $\re f(0)=0$ and $R_I(0)=[\nu(0)/\nuF]^2$. For the transverse channel the same construction with the $+$ sign gives $\int(-\partial_\ee n_F)[(\re g)^2+(\re f)^2]$, whose logarithmic divergence at the gap edge (cut off by $\omega$) is the origin of the coherence peak in Eq.~\eqref{eq:HSasympt}; in the full kernel \eqref{eq:kernelfinal} this structure is contained in the coherence factors \eqref{eq:cohfactors}.


\section{Numerical details}
\label{app:numerics}

All results were produced with a Python implementation of the formalism, structured as follows.

\emph{Shiba solution.}  Equations~\eqref{eq:selfenergies} are solved on a graded energy grid (clustered at the gap edges) by fixed-point iteration with Aitken extrapolation, selecting the retarded branch by $\im D\ge0$ and the physical sign by $\re g\ge0$; the converged $D_s=\sqrt{\tilde\ee_s^2-\tilde\Delta_s^2}$ agrees with Eq.~\eqref{eq:hcorrect} to $10^{-10}$. For $1/\tau_s=0$ the closed-form BCS expressions are used, since spline interpolation of the $1/\sqrt{\ee^2-\Delta^2}$ singularities is numerically unreliable.

\emph{Gap.}  Equations~\eqref{eq:shibamatsu}--\eqref{eq:gapeq} are solved by vectorized Newton iteration in $u_n$ and bisection in $\Delta$, with a few thousand Matsubara frequencies and an analytic tail correction.

\emph{Kernel.}  The energy integral in Eq.~\eqref{eq:kernelfinal} is evaluated at the physical $\omega$ by Gauss--Legendre quadrature on panels between breakpoints placed at $\pm\Delta\pm\omega/2$ (with geometric refinement down to $10^{-6}\Delta$), at $\pm\omega/2$, and at the thermal scales; the function $\mathcal A$ is evaluated by its large-$l_0$ series for $|l_0|>30$ to avoid cancellation. Keeping the full $\omega$ dependence (rather than the low-frequency expansion) is essential for $1/\tau_s=0$, where the coherence peak is cut off by $\omega$ [Eq.~\eqref{eq:HSasympt}].

\emph{Surface and noise integrals.}  $\eps_t(q,\omega)$ is evaluated on a logarithmic $q$ grid and interpolated; $\zeta_{s,p}(u)$ are computed by quadrature after subtracting the vacuum ($\eps\to1$) integrands, whose integrals are known analytically; the $w$ integrals \eqref{eq:bpar}--\eqref{eq:bperp} are performed on grids scaled to $1/2kz$.

\emph{Validation.}  The implementation was checked against independent results: the exact residue evaluation of the $\xi$ integrals \eqref{eq:xiRA}--\eqref{eq:xiRR} (symbolically, after $b$-symmetrization); the Fresnel coefficients in the local limit (to $10^{-6}$); the Drude limit of Eq.~\eqref{eq:mermin2} (symbolically); the clean BCS gap, $\Delta_0/T_{c0}=1.7639$ and $\Delta(0.5T_c)/\Delta_0=0.9569$; the Abrikosov--Gor'kov critical pair-breaking rate; the normal-state kernel \eqref{eq:QN} for $\Delta\to0$ (agreement to better than $3\%$ over $q=10^4$--$10^8$~m$^{-1}$); the London limit $\hat Q\to1$ (to $10^{-3}$); the Pippard limit \eqref{eq:pippard}; the Mattis--Bardeen conductivity \eqref{eq:MB} (to $10^{-3}$ for $0.3\le T/T_c\le0.95$); the Shiba density of states of Ref.~\cite{kharitonov}, Fig.~1; the residual conductivity $\sigma_1(0)/\sigma_n=[\nu(0)/\nuF]^2$; and the $b$ integrals against independent adaptive quadrature (to $2\%$). All figures in this paper were generated from this validated code.

\bibliography{noise_sc_fluct}

\end{document}